\documentclass[sigconf]{acmart}
\renewcommand\footnotetextcopyrightpermission[1]{} % removes footnote with conference information in first column
\usepackage{graphicx}
\usepackage{balance}
\usepackage{url}
\usepackage{amsmath}
\usepackage{color,soul}
\usepackage[linesnumbered,ruled,vlined]{algorithm2e}
\usepackage{dsfont} %\mathds{}
\usepackage{amssymb} %\mathbb{}
\usepackage{mathrsfs} %\mathscr{}
\usepackage{colortbl} % use colortbl to replace xcolor.
\usepackage{multirow}
\usepackage{bm}
\usepackage{amssymb}% http://ctan.org/pkg/amssymb
\usepackage{pifont}% http://ctan.org/pkg/pifont
\usepackage{array}%to define L for table word wrap
\usepackage{tablefootnote}

\newcolumntype{B}{>{\centering\arraybackslash}m{2.3cm}}
\newcolumntype{D}{>{\centering\arraybackslash}m{3.3cm}}
\newcolumntype{E}{>{\centering\arraybackslash}m{2.1cm}}
\newcommand{\cmark}{\ding{51}}%
\newcommand{\xmark}{\ding{55}}%

\newcommand{\oldcomments}[1]{}   % make the comments private

\newcommand{\newcomments}[1]{}   % make the comments private

\newcommand{\yannisp}[1]{\oldcomments{\textcolor{blue}{YannisP: #1}}}

\newcommand{\rone}[1]{\newcomments{\textcolor{red}{[[[Review 1]]]: #1}}}
\newcommand{\rtwo}[1]{\newcomments{\textcolor{blue}{[[[Review 2]]]: #1}}}
\newcommand{\rthree}[1]{\newcomments{\textcolor{green}{[[[Review 3]]]: #1}}}
\newcommand{\eat}[1]{}

\newcommand{\db}{Plato}
\newcommand{\cover}[2]{\Pi_{#1,#2}}
	
\newcommand{\ts}{\bm{T}}
\newcommand{\tsi}[1]{\bm{T}_{#1}}
\newcommand{\ser}{\textsf{Constant}}
\newcommand{\shift}{\textsf{Shift}}

\newcommand{\timesOP}{\bm{T_1\times T_2}}
\newcommand{\plusOP}{\bm{T_1+T_2}}
\newcommand{\minusOP}{\bm{T_1-T_2}}
\newcommand{\ff}{\mathds{F}} % function family
\newcommand{\ef}[1]{f^*_{#1}} % estimation function of T_i^j
\newcommand{\es}[1]{\Phi(#1)} % error measures of T_i^j
\newcommand{\fes}[1]{\|\varepsilon_{#1}\|_2} % first error measure of T_i^j
\newcommand{\ses}[1]{\|f_{#1}\|_2} % second error measure of T_i^j
\newcommand{\tes}[1]{\gamma_{#1}} % third error measure of T_i^j
\newcommand{\te}{\varepsilon}
\newcommand{\eg}[1]{\hat{\varepsilon}_{#1}}
\newcommand{\rep}[1]{\tilde{#1}}

\newtheorem{theorem}{Theorem}

\newtheorem{definition}{Definition}
\newtheorem{example}{Example}

\newenvironment{compact_item}
{\setlength{\leftmargini}{2em}
\begin{itemize}
  \setlength{\labelsep}{1em}
  \setlength{\itemsep}{.1em}
  \setlength{\parskip}{0pt}
  \setlength{\parsep}{0pt}}
{\end{itemize}}

\newtheorem{lemma}{Lemma}

\begin{document}
\title{Plato: Approximate Analytics over Compressed Time Series with Tight Deterministic Error Guarantees}
%\title{Plato: Estimating Queries over Compressed Time Series with Tight Deterministic Error Guarantees}
%\title{Plato: Query Estimation with Tight Deterministic Error Guarantees}

\author{Chunbin Lin}
\affiliation{%
  \institution{University of California, San Diego}
}
\email{chunbinlin@cs.ucsd.edu}

\author{Etienne Boursier}
\affiliation{%
  \institution{ENS Paris-Saclay}
}
\email{eboursie@ens-paris-saclay.fr}

\author{Yannis Papakonstantinou}
\affiliation{%
  \institution{University of California, San Diego}
}
\email{yannis@cs.ucsd.edu}

\begin{abstract}
\db\ provides fast approximate analytics on time series, by precomputing and storing compressed time series. \db's key novelty is the delivery of \textit{tight deterministic error guarantees} for time series analytics. \db{} evaluates any time series expression composed by the linear algebra operators over vectors, along with arithmetic operators. This large scope of possible expressions includes common use cases such as correlation and cross-correlation expressions.  Each time series is segmented either by fixed-length segmentation or by (a usually more effective) variable-length segmentation. Each segment is compressed by an estimation/compression function that approximates the actual values and is coming from a user-chosen function family, as taught by many prior works. The novelty is that \db\ associates to each segment 1 to 3 (depending on the case) precomputed error measures and, using them, \db\ computes tight deterministic error guarantees for analytics over the compressions.

%While \db\ allows applying any of the well-known estimation/compression function families (e.g., polynomial, gaussians, etc) to compress time series, 
Importantly, some compression families lead to much better deterministic error guarantees. This work identifies two broad estimation function family groups (\textit{Vector Space (VS)} and \textit{Linear Scalable Family (LSF)}),  which lead to theoretically and practically high-quality guarantees, even for expressions (eg correlation) that combine multiple time series that have been independently compressed and may, thus, use misaligned segmentations. 
The theoretical aspect of ``high quality'' is crisply captured by the \textit{Amplitude Independence (AI)} property: An AI guarantee does not depend on the amplitude of the involved time series, even when we combine multiple time series. The experiments on four real-life datasets showed that when the novel AI guarantees were applicable, the approximate query results were certified to be very close (typically $1\%$) to the true results.
%In summary, this work shows that deterministic error guarantees are feasible and practical, given the appropriate combination of error measures and estimation function family.
\end{abstract}

\maketitle

\section{Introduction}
\label{sec:introduction}
%Motivation

\begin{figure}[t]
\center
\includegraphics[width=0.5\textwidth]{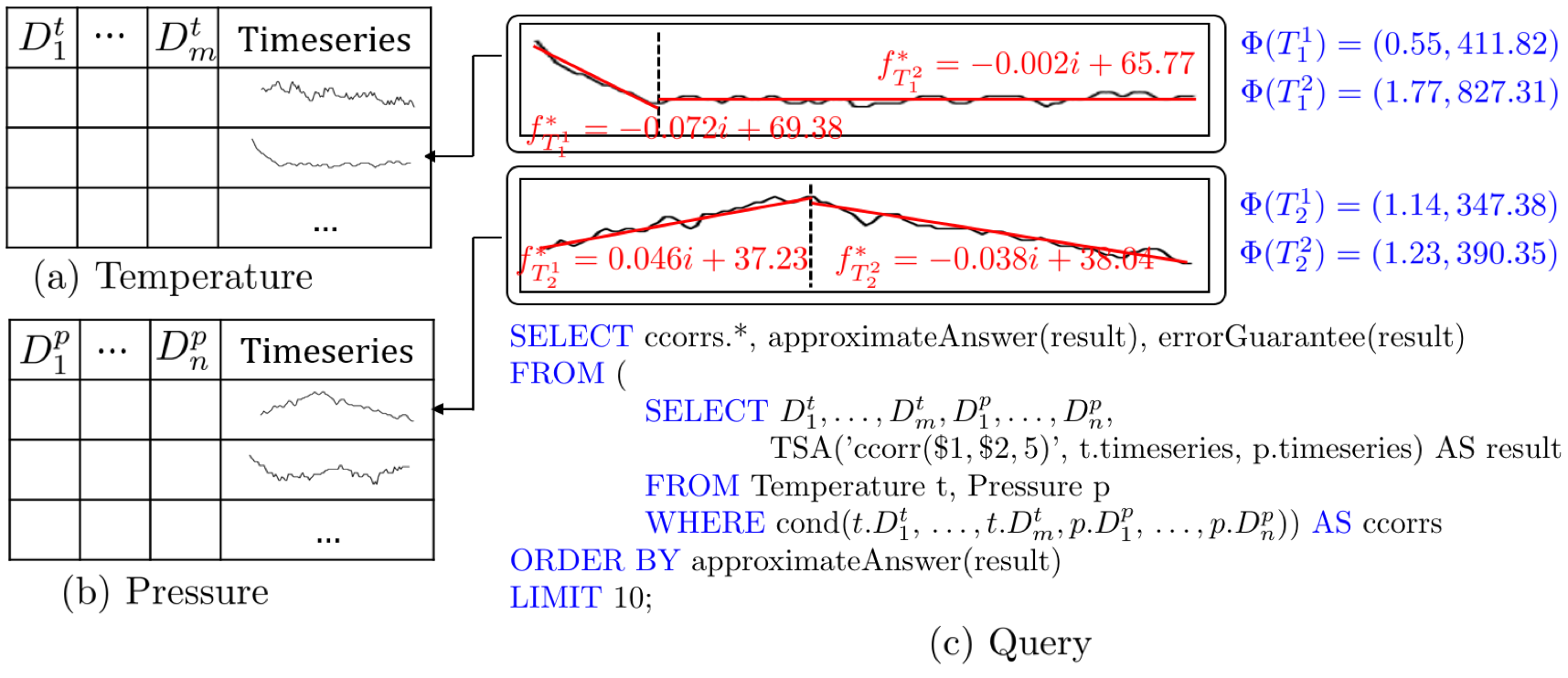}
\caption{Example of SQL query using the time series analytic (TSA) UDF.}
\label{fig:intro_time}
\end{figure}

%Analytics over historical time series data is common in a variety of domains including finance, transportation, public health, environmental protection, and failure prediction. 
Attention to time series analytics is bound to increase in the IoT era as cheap sensors can now deliver vast volumes of many types of measurements. The size of the data is also bound to increase. E.g., an IoT-ready oil drilling rig produces about $8$ TB of operational data in one day.~\footnote{https://wasabi.com/storage-solutions/internet-of-things/} One way to solve this problem is to increase the expense in computing and storage in order to catch up. However, in many domains, the data size increase is expected to outpace the increase of computing abilities, thus making this approach unattractive~\cite{galakatos2017revisiting,chaudhuri2017approximate}. Another solution is \textit{approximate analytics} over compressed time series.

\begin{table*}[t]
\centering
\renewcommand{\tabcolsep}{2mm}
\begin{tabular}{|c|c|c|c|c|c|c|c|}
\hline
\multirow{2}{*}{}                          &function                           & error guarantees on                                                            & \multirow{2}{*}{AI}       & \multirow{2}{*}{Tight}  & error guarantees on                                                                                                                           & \multirow{2}{*}{AI}     & \multirow{2}{*}{Tight}   \\
                                           &family                             & aligned time series                                                            &                           &                         & misaligned time series                                                                                                                        &                         &                          \\\hline
\multirow{8}{*}{$Sum(\bm{T_1\times T_2})$} & \multirow{3}{*}{ANY$\setminus$VS} & $\sum\limits_{i=1}^k\Big(\fes{T_1^i}\times \fes{T_2^i}\Big)$                   & \multirow{3}{*}{\xmark}   & \multirow{3}{*}{\cmark} & $\;\;\sum_{i=1}^{k_1} \Big(\fes{T_1^i}\times(\sum_{j\in \cover{T_2}{[a_1^i,b_1^i]}}\ses{T_2^j}^2)^{\frac{1}{2}}\Big)$                             &\multirow{4}{*}{\xmark}  & \multirow{4}{*}{\cmark}  \\
                                           &                                   & $+ \sum\limits_{i=1}^k\Big(\fes{T_1^i}\times \ses{T_2^i}\Big)$                 &                           &                         & $+\sum_{i=1}^{k_2} \Big(\fes{T_2^i}\times(\sum_{j\in \cover{T_1}{[a_2^i,b_2^i]}}\ses{T_1^j}^2)^{\frac{1}{2}}\Big)$                            &                         &                          \\
                                           &                                   & $+ \sum\limits_{i=1}^k\Big(\fes{T_2^i}\times \ses{T_1^i}\Big)$                 &                           &                         & $+\sum\limits_{[a,b] \in OPT(L_{T_1}, L_{T_2})} \Big(\big(\sum_{i\in \cover{T_1}{[a,b]}} \|\varepsilon_{T_1^i}\|_2^2\big)^{\frac{1}{2}}$ &                         &                          \\\cline{2-5}
                                           & VS$\setminus$LSF                  &  \multirow{5}{*}{$\sum\limits_{i=1}^k \Big(\fes{T_1^i}\times \ses{T_2^i}\Big)$}& \multirow{5}{*}{\cmark}   & \multirow{5}{*}{\cmark} & $\;\;\;\;\;\;\;\;\;\;\;\;\;\;\;\;\;\;\;\;\;\;\;\times \big(\sum_{i\in \cover{T_2}{[a,b]}} \|\varepsilon_{T_2^i}\|_2^2\big)^{\frac{1}{2}}\Big)$                                        &                         &                          \\\cline{2-2}\cline{6-8}
                                           &\multirow{4}{*}{LSF}               &                                                                                &                           &                         & $\;\;\sum_{i=1}^{k_1}\Big(\fes{T_1^i}\times \|\bm{f_{T_2}|_{[a_1^i,b_1^i]} - f^*_{T_1^i}}\|_2\Big)$                                               &\multirow{4}{*}{\cmark}  & \multirow{3}{*}{\cmark}  \\
                                           &                                   &                                                                                &                           &                         & $+\sum_{i=1}^{k_2}\Big(\fes{T_2^i}\times \|\bm{f_{T_1}|_{[a_2^i,b_2^i]} - f^*_{T_2^i}}\|_2\Big)$                                              &                         &                          \\
                                           &                                   &                                                                                &                           &                         & $+\sum\limits_{[a,b] \in OPT(L_{T_1}, L_{T_2})} \Big(\big(\sum_{i\in \cover{T_1}{[a,b]}} \|\varepsilon_{T_1^i}\|_2^2\big)^{\frac{1}{2}}$ &                         &                          \\
                                           &                                   &                                                                                &                           &                         & $\;\;\;\;\;\;\;\;\;\;\;\;\;\;\;\;\;\;\;\;\;\;\;\times \big(\sum_{i\in \cover{T_2}{[a,b]}} \|\varepsilon_{T_2^i}\|_2^2\big)^{\frac{1}{2}}\Big)$                                         &                         &                          \\\hline
\multirow{2}{*}{$Sum(\bm{T_1+T_2})$}       &\multirow{4}{*}{ANY}               &\multirow{4}{*}{$\sum\limits_{i=1}^k(\tes{T_1^i}+\tes{T_2^i})$}                 &\multirow{4}{*}{\cmark}    &\multirow{4}{*}{\cmark}  & \multirow{4}{*}{$\sum\limits_{i=1}^{k_1}\tes{T_1^i}+\sum\limits_{j=1}^{k_2}\tes{T_2^j}$}                                                      &\multirow{4}{*}{\cmark}  & \multirow{4}{*}{\cmark}  \\
                                           &                                   &                                                                                &                           &                         &                                                                                                                                               &                         &                       \\\cline{1-1}
\multirow{2}{*}{$Sum(\bm{T_1-T_2})$}       &                                   &                                                                                &                           &                         &                                                                                                                                               &                         &                       \\
                                           &                                   &                                                                                &                           &                         &                                                                                                                                               &                         &                       \\\hline
\end{tabular}
\caption{Error guarantees for the time series analytic (TSA) $Sum(T_1\diamond T_2)$ where $\diamond\in \{\times, +, -\}$ on both aligned and misaligned time series compressed by estimation functions in different families. We assume  $T_1$ and $T_2$ have $k_1$ and $k_2$ segments respectively. In the aligned case, we have $k_1=k_2=k$. $OPT(L_{T_1}, L_{T_2})$ is the optimal segment combination returned by the algorithm OS in Section~\ref{sec:segment_combination_selection}}
\label{tab:error_estimation}
\end{table*}

%Approximate analytics enables fast computation over historical time series data. 
\begin{example}
Consider the database in Figure~\ref{fig:intro_time}, which has a \textsf{Temperature} table and a \textsf{Pressure} table. Each table contains (i) one \textsf{Timeseries} column containing time series data, as a UDT \cite{eisenberg2002sql} and (ii) several other ``dimension" attributes $D$, such as the identification and other properties of the sensors that delivered the time series. The \db{}\ SQL query in Figure~\ref{fig:intro_time}(c) ``returns the top-10 temperature/pressure 5-second cross-correlation scores among all the (temperature, pressure) pairs satisfying a (not detailed in the example) condition \textit{cond()} over the dimension attributes''. The \textit{Time Series Analytic} (\textsf{TSA}) function evaluates the time series expression \textsf{'CCorr(t.timeseries, p.timeseries, 5)'}. (The \textsf{\$1} is instantianted by the second argument of the \textsf{TSA} and similarly for the \textsf{\$2}.) In SQL, the \textsf{result} is a string concatenation of the approximate answer and the error guarantee. The functions \textsf{approximateAnswer} and \textsf{errorGuarantee} extract the respective pieces. The filtering conditions in the WHERE clause have no effects to our techniques as a time series is either completely selected or completely pruned. For the case where only a part of the time series is chosen, see the \textsf{Restriction} operator in Section~\ref{sec:DataQueries}. Furthermore, this work is orthogonal to how to assemble the time series values from plain relational tables that have no time series type.

Note, instead of using the cross-correlation expression \textsf{CCorr} the user could use more basic functions to write out the definition of cross-correlation, as shown in Table~\ref{tbl:common-statistics-queries}. The result would be the same. Either way, computing the accurate cross-correlations would cost more than $10$ minutes, in the experimental setting of Figure~\ref{fig:align_misalign_time}, in Experiments. However, \db{} reduces the runtime to within one second by computing the approximate correlations. It also delivers deterministic error guarantees, which means the error guarantees have a $100\%$ confidence. 
\end{example}

\noindent\textbf{Deterministic error guarantees vs. probabilistic error guarantees.} Both deterministic error guarantees~\cite{DBLP:conf/icde/CormodeKMS05,DBLP:conf/sigmod/GreenwaldK01,DBLP:conf/sigmod/RajagopalanML98,PottiP15,LazaridisM01,PoosalaIHS96} and probabilistic error guarantees~\cite{chaudhuri2007optimized,SidirourgosKB11,PansareBJC11, AgarwalMPMMS13} are widely utilized in approximate query processing. We follow the deterministic error guarantees direction. Section~\ref{subsec:experiment} compares deterministic and probabilistic guarantees.

The success of approximate querying on IoT time series data is based on an important beneficial property of time series data: the points in the sequence of values normally \textit{depend} on the previous points and exhibit \textit{continuity}.
For example, a temperature sensor is very unlikely to report a 100 degrees increase within a second. Therefore, in the signal processing and data mining communities~\cite{KeoghCPM01,keogh1997fast,FaloutsosRM94,ChanF99}, time series data is usually modeled and compressed by continuous functions in order to reduce its size. For instance, the Piecewise Aggregate Approximation (PAA)~\cite{KeoghCPM01} and the Piecewise Linear Representation (PLR)~\cite{keogh1997fast} adopt polynomial functions (0-degree in PAA and 1-degree in PLR) to compress the time series; \cite{pan2017construction} uses Gaussian functions; \cite{tobita2016combined} applies natural logarithmic functions and natural exponential functions to compress time series. \db{} is open to any existing time series compression technique, since there is no one-size-fits-all compression function family that can best model all kinds of time series data. For example, polynomials and ARMA models are better at modeling data from physical processes such as temperature \cite{MeiM17,choi2012arma}, while Gaussian functions are better for modeling relatively randomized data \cite{Kim03} such as stock prices. 
%In addition, different applications prefer different function families. For example, linear adaptive algorithms (e.g., LMS~\cite{borisagar2010simulation}) prefer polynomial functions, while nonlinear filtering (e.g., Kalman Filter~\cite{ArulampalamMGC02}) uses Gaussian functions.~\footnote{http://www.iva.cs.tut.fi/HomePage/current.html} 
How to choose the best compression function family has been widely studied in prior work~\cite{philo1997improved,wiscombe1977exponential,denison1998automatic,kovacs2002box} and recent efforts even attempt to automate the process \cite{KumarMNP15}. In this paper, the choice of compression function family is made by the user. \db\ will deliver the best guarantees, given the chosen family.

\begin{figure}[t]
\center
\includegraphics[width=0.46\textwidth]{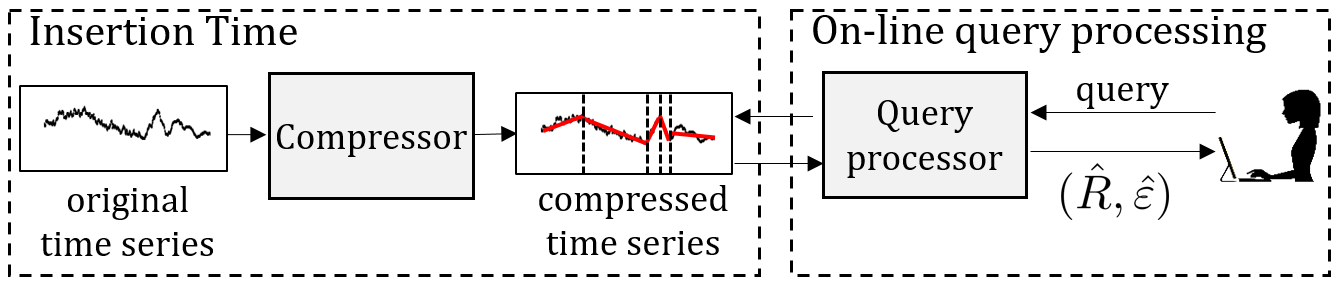}
\vspace{-3mm}
\caption{\db's Approximate Querying}
\label{fig:basic-arch}
\end{figure}

% Define architecture of approximate querying, show that the technique works
\noindent\textbf{Architecture}. 
Figure~\ref{fig:basic-arch} shows the high-level architecture. During insertion time, the provided time series is compressed. In particular, a compression \textit{function family} (e.g., 2nd-degree polynomials) is chosen by the user.
Internally, in a simple version, each time series is segmented (partitioned) first in equal lengths. Then, for each segment the system finds the best \textit{estimation function}, which is the member of the function family that best approximates the values in this segment. The most common definition of ``best'' is the minimization of the \textit{reconstruction error}, i.e., the minimization of the Euclidean distance between the original and the estimated values. This is also the definition that \db{} assumes. The compressed database stores the parameters of the estimation function for each segment, which take much less space than the original time series data. In the  more sophisticated version, segmentation and estimation are mingled together \cite{koski1995syntactic,keogh2001online} to achieve better compression. The result is that the time series is partitioned into variable-length segments.

Consequently, given a query $q$ with \textit{Time Series Analytics} (\textsf{TSA}) UDF calls,
the database computes quickly an approximate answer for each \textsf{TSA} call by using the compressed data. 
%Note, TSAs may combine multiple time series; e.g., this is the case with correlation and cross-correlation.

\begin{example}
\label{xmpl:intro-demo}
Consider a room temperature time series $T_1$ and an air pressure time series $T_2$ in Figure~\ref{fig:intro_time} and consider the \textsf{TSA}(`Ccorr($T_1, T_2$, $60$)', $T_1$, $T_2$) where `Ccorr($T_1, T_2, 60$)' refers to the 60-seconds cross-correlation of $T_1$ and $T_2$ (see definition in Table~\ref{tbl:common-statistics-queries}). Both $T_1$ and $T_2$  have $600$ data points at 1-second resolution and are segmented by variable length segmentation methods and compressed by PLR (1-degree polynomial functions). 
%Instead of storing the original data, \db{} compresses the time series. Assume both $T_1$ and $T_2$ are compressed by PLR, i.e., . Notice that the variable length segmentation leads to much better estimation/approximation (lower reconstruction errors) than what we would get by splitting each time series into two equal segments.
\eat{
Assume an analyst wants to compute the 1-minute cross-correlation between $T_1$ and $T_2$, which is expressed by the time series expression
$(\sum_{i=1}^{600}((d^t_i-\mu_{\mathbf{t}})(d^p_{i+60}-\mu_{\mathbf{p}})))/(\sigma_{\mathbf{t}}\times \sigma_{\mathbf{p}})$.}
The precise answer is  $0.303$. But instead of accessing the $1200$ ($600\times2$) original data points, \db{} produces the approximate answer $0.300$ (error is $0.003$) by accessing just the function parameters ($-0.072,69.38$), ($-0.002, 65.77$) for $T_1$ and ($-0.046,37.23$), ($-0.038, 38.04$) for $T_2$ in the compressed database.%
\footnote{Due to reasons relating to computation efficiency, as explained in Section~\ref{sec:error_guarantee_misalign_LSF}, \db{} does not actually store the parameters ($-0.072,69.38$), ($-0.002, 65.77$) and ($-0.046,37.23$), ($-0.038, 38.04$) in their standard basis but rather it stores coefficients in an orthonormal basis.}
\end{example}

% What this paper is not about

The well-known downside of approximate querying is that errors are introduced. When the example's user receives the approximate answer $0.300$ she cannot tell how far this answer is from the \textit{true answer}, i.e., the precise answer. The novelty of \db{} is the provision of \textit{tight (i.e., lower bound) deterministic error guarantees} for the answers, even when the time series expressions combine multiple series. 
%,DBLP:conf/icde/ZhangLXKW06
%, which is orthogonal to techniques providing probabilistic error guarantees.
In the Example~\ref{xmpl:intro-demo}, \db{} guarantees that the true answer is within $\pm 0.0032$ of the approximate answer $0.300$ with $100\%$ confidence. (Indeed, $0.303$ is within $\pm 0.0032$ of $0.300$.)
% It produces these guarantees by utilizing \textit{error measures} associated with each segment.

\noindent \textbf{Scope of Time Series Expressions}. \db{} supports the time series analytic expressions formally defined in Table~\ref{table:query} (Section~\ref{sec:DataQueries}). They are composed of vector operators ($\mathbf{+}$, $\mathbf{-}$, $\mathbf{\times}$, \textsf{Shift}), arithmetic operators, the aggregation operator \textsf{Sum} that turns its input vector into a scalar, and the \textsf{Constant} operator that turns its input scalar into a vector. As such, \db{} queries can express not only statistics that involve one time series (eg, average, variance, and n-th moment) but also  statistics that involve multiple time series, such as correlation and cross-correlation.

\yannisp{Many problems on the table of example statistics: 1. Add the definition of $\mu$ and $\sigma$ in the table (as queries, on the top), above the Covariance so that the $\mu$ and $\sigma$ do not come out of nowhere. 2. The way you compute the query expressions does not correspond with the definition. In the case of covariance, use the Serialize operator to make vectors out of the $\mu$. That is $\frac{1}{n} \times Sum((\bm{T_1} - Serialize(\mu_{T_1})) \bm{times} (\bm{T_2} - Serialize(\mu_{T_2})))$. Same for the next expressions. This has the advantage that you exhibit the Serialize and you make an obvious correspondence between definition and query expression. (Not one that would require an algebra book to figure out.)
3. Since they will be too many, eliminate the covariance, cross-covariance and auto-covariance. They are obviously doable once correlation, cross-correlation are doable.
4. Shorten the caption to have only the first sentence. What's the $d$'s anyway? They are never used. Chunbin: fixed.
}

\noindent \textbf{Goal of Plato}.
The goal of Plato is to provide tight deterministic error guarantees for the time series expressions.  It is challenging as each time series is segmented and compressed individually before the queries arrive, which results in (i) time series segmentations are misaligned, and (ii) different compression functions are utilized in different time series.
To solve the challenge, Plato computes error measures of Table~\ref{tab:error_measures} during the compression time for each time series segment.
Figure~\ref{fig:intro_time} shows the error measures $\Phi$ (in blue) for each segment of the example. With the help of the error measures, no matter whether a time series is compressed by trigonometric functions or polynomial functions or some other family, \db{} is able to give tight deterministic error guarantees for queries involving the compressed time series.

{\small
\begin{table}[t]
\centering
\renewcommand{\tabcolsep}{1.3mm}
\begin{tabular}{|l|l|}
  \hline
Error measures & Comments\\\hline
$\fes{T} =  \sqrt{\sum_{i=a}^b (T[i]-\ef{T}(i))^2}$ & $L_2$-norm of the estimation errors\\\hline
$\ses{T} =  \sqrt{\sum_{i=a}^b (\ef{T}(i))^2}$ & $L_2$-norm of the estimated values\\\hline
$\tes{T} = |\sum_{i=a}^b T[i]  - \sum_{i=a}^b \ef{T}(i)|$ & Absolute reconstruction error\\\hline
\end{tabular}
\caption{Error measures stored for a time series segment $T$ running from $a$ to $b$ and approximated with the estimation function $\ef{T}$.}
\label{tab:error_measures}
\end{table}
}
\normalsize
\yannisp{Just place in the figures the three error measures per segment. Saves space. Increases readability. No need to type the computation. Just the results. Be careful with the typesetting and fonts to look similar to the text. Chunbin: done.}
%The error measures for the first temperature segment $\mathbf{t_1}$, running from $1$\yannisp{complete. Chunbin:done} to $238$\yannisp{complete. Chunbin:done} are computed as follows: $\|\mathbf{\varepsilon}_{\mathbf{t_1}} \|_2 =  (\sum_{i=1}^{238} |\mathbf{t}[i]  - (-0.072i+69.38)|^2)^{\frac{1}{2}} = 1.05$, $\xi_{\mathbf{t_1}} = |\sum_{i=1}^{238} \mathbf{t}[i] - \sum_{i=1}^{238} (0.035i+51.16)|= 0$, $\|\hat{f}^{\mathbf{t_1}}\|_2 =  (\sum_{i=1}^{238} (-0.072i+69.38)^2)^{\frac{1}{2}} = 940.71$.
%Notice, the measure $\ses{T}$ is not really about errors. Indeed, it may not need to be stored, since it can be often computed from the estimation function's parameters, using fast symbolic computation.
%This is the case when estimating with polynomials, as in~Example~\ref{xmpl:intro-demo}. Thus, the error measures truly stored for, say, the segment $T_1^1$ are just $(0.55, 411.82, 0)$.~\footnote{We will show in Section~\ref{subsec:error_statistics} that the last error measure usually can be avoided as it is usually guaranteed to be 0.}

\noindent \textbf{Function Family Groups Producing Practical Error Guarantees}.
\db{} produces tight error guarantees, for \textit{any} function family that may have been used in the compression. In addition, our theoretical and experimental analysis identifies which families lead to high quality guarantees.

The formulas of Table~\ref{tab:error_estimation} provide error guarantees for characteristic, simple expressions and exhibit the difference in guarantee quality. Any other expression, e.g., the statistics of Table~\ref{tbl:common-statistics-queries}, are also given error guarantees by composing the error measures and guarantees of their subexpressions (as shown in the paper) and the same quality characterizations apply to them inductively.

\begin{figure}[t]
\center
\includegraphics[width=0.49\textwidth]{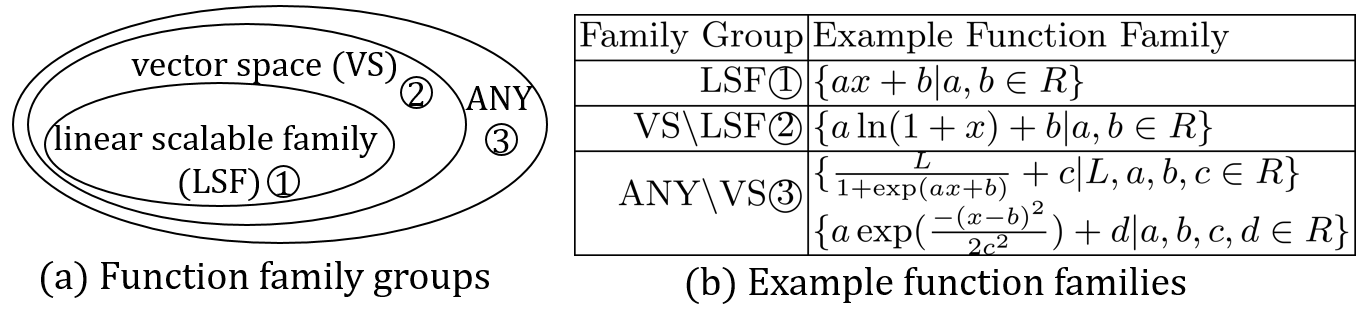}
\vspace{-7mm}
\caption{Function family groups and examples.}
\label{fig:function_family_xmpls}
%\center
%\includegraphics[width=0.5\textwidth]{Figure/dichotomy.png}
%\vspace{-7mm}
%\caption{Function family groups and resulting guarantees}
%\label{fig:dichotomy}
\end{figure}

This is how to interpret the results of Table~\ref{tab:error_estimation}: Three function family groups have been identified: (1) The Linear Scalable Family group (LSF), (2) the Vector Space (VS), which includes the LSF and (3) ANY, which, according to its name, includes everything. Given the function family $\mathds{F}$ used in the compression, we first categorize $\mathds{F}$ in one of LSF or VS/LSF (i.e., VS excluding LSF) or ANY/VS. For example, if $\mathds{F}$ is the 2-degree polynomials, then $\mathds{F}$ belongs to LSF. See Figure~\ref{fig:function_family_xmpls} for other examples. \yannisp{I moved the table here. Remove it from later. Chunbin: done.} Next, we consider whether the segments of the involved compressed time series are aligned or misaligned and finally we look at the error guarantee formula for the expression.

\noindent\textbf{Amplitude Independence (AI)}.
The specifics of interpreting the table's results and the specifics of their efficient computation require the detailed discussion of the paper. 
%(Eg, the summation index $OPT$ corresponds to the optimal segment combination (Section~\ref{sec:segment_combination_selection}.) 
%(Eg, the summation index in some formulas corresponds to a non-trivial algorithm that chooses segments.) 
Nevertheless, a  clear and general high level lesson about the practicality of the error guarantees emerges from the table's summary: \textit{Some function families allow for much higher quality error guarantees than other function families.} The typical characteristic of ``higher quality'' is \textit{Amplitude Independence (AI)}. If an error guarantee is AI, then it is not influenced by the $\ses{T}$ measure, i.e., it is not affected by the amplitude of the values of the estimation functions and, thus, it is not affected from the amplitude of the original data. An AI error guarantee is only affected by the reconstruction errors caused by the estimation functions, which intuitively implies that AI error guarantees are close to the actual error.

\noindent\textbf{Tight Error Guarantees }.
These guarantees are \textit{tight} in the following sense. Given (a) the function family categorization  into LSF, VS/LSF or ANY/VS and (b) segments with the error measures of Table~\ref{tab:error_measures}, the formula provided by Table~\ref{tab:error_estimation} produces an error guarantee that is as small as possible. That is, for this superfamily and for the given error measures, any attempt to create a better (i.e., smaller) error guarantee will fail because there are provably time series and at least one time series analytics expression where the true error is exactly as large as the error guarantee.

The experimental results, where we tried data sets with different characteristics and different compression methods, verified the above intuition: AI error guarantees were \textit{order(s) of magnitude smaller} than their amplitude dependent counterparts. 
%Indeed, AI ones over variable-length compressions were invariably small enough to be practically meaningful, while non-AI guarantees were too large to be practically useful. 
% Looking at Table~\ref{tab:error_estimation} we see that queries that involve only addition and subtraction of vectors produce AI guarantees no matter what function family is used. This is not surprising, given the linearity of addition and subtraction.
Particularly interesting are the analytics that combine multiple vectors by vector multiplication (eg, cross-correlation). Then the amplitude independence of the error guarantees does not apply to all families and, thus, this study says that one should be careful in the choice, if error guarantees are needed. 
The contributions are summarized as follows.
\begin{compact_item}
\item We deliver tight deterministic error guarantees for a wide class of analytics over compressed time series. The key challenge is analytics (e.g., correlation and cross-correlation) that combine multiple time series but it is not known in advance which time series may be combined. Thus, each time series has been compressed individually, much before a query arrives. The reconstruction errors of the individual time series' compressions cannot provide, by themselves,  decent guarantees for queries that multiply time series. To make the problem harder, time series segmentations are generally misaligned.%
\footnote{Misalignment happens because the most effective compressions use variable length segmentations. But even if the segmentations were fixed length, queries such as cross-corellation and cross-autocorellation time shift one of their time series, thus producing misalignment with the second time series.} \yannisp{Notice that the footnote motivates misalignment. Thus you can remove the misalignment motivation from the subsequent section. Chunbin: I see. }
\item The provided guarantees apply regardless of the specifics of the segmentation and estimation function family used during the compression, thus making the provided deterministic error guarantees applicable to any prior work on segment-based compression (eg, variable-sized histograms etc). The only requirement is the common assumption that the estimation function minimizes the Euclidean distance between the actual values and the estimates.
\item  We  identify broad estimation function family groups (namely, the already defined Vector Space family and the presently defined Linear Scalable Family) that lead to theoretically and practically high quality guarantees. The theoretical aspect of high quality is crisply captured by the Amplitude Independence (AI) property.  Furthermore, the error guarantees are computed very efficiently, in time proportional to the number of segments.
\item The results broadly apply to analytics involving composition of the typical operators, which is powerful enough to express common statistics, such as variance, correlation, cross-correlation and other in any time range.
\item We conduct an extensive empirical evaluation on four real-life datasets to evaluate the error guarantees provided by \db{} and the importance of the VS and LSF properties on error estimation. The results show that the AI error guarantees are very narrow - thus, practical. Furthermore, we compare to sampling-based approximation and show experimentally that \db{} delivers deterministic (100\% confidence) error guarantees using fewer data than it takes to produce probabilistic error guarantees with 95\% and 99\% confidence via sampling.
%
% \item Section~\ref{sec:summary_future} lists the many future directions that can be followed from here on.
\end{compact_item}

\eat{
\noindent\textbf{Paper organization} Time series and expressions are defined in Section~\ref{sec:DataQueries}. Section~\ref{sec:time_series_index} introduces the compressed segment index structure. The computation of error guarantees is proposed in Section~\ref{sec:error_estimation}. Experiment results are presented in Section~\ref{sec:experiments}. %Section~\ref{sec:related_work} presents related work and 
Section~\ref{sec:summary_future} provides the conclusion and future directions. The related work, proofs and additional experiments are in the Appendix.
% Section~\ref{appendix:proof_of_polynomial_LSF} is the Appendix section. 
}
\section{Time Series and Expressions}
\label{sec:DataQueries}

\rthree{The ``queries'' considered in the paper are rather limited statistics, though authors argue that other more complex analytics queries can also be implemented using the considered building blocks.}

{\small
\begin{table}[t]
\centering
\renewcommand{\tabcolsep}{0.5mm}
\begin{tabular}{|cll|c|}
\hline
\multicolumn{4}{|l|}{\cellcolor{gray!15} \text{Time Series Analytic (TSA)}}\\
\hline
Q & $\rightarrow$ & Ar                      &\\
\hline
%\multicolumn{4}{l}{}\\
\hline
\multicolumn{4}{|l|}{\cellcolor{gray!15} \text{Arithmetic Expression (Ar)}}\\
\hline
Ar & $\rightarrow$ & \text{literal value in $R$} & \\
			   & |           & Ar $\otimes$ Ar & \text{where } $\otimes\in\{+,-,\times,\div,\sqrt{~}\}$\\
         & |           & Agg	&\\
\hline
%\multicolumn{4}{l}{}\\
\hline
\multicolumn{4}{|l|}{\cellcolor{gray!15} \text{Aggregation Expression (Agg)}}\\
\hline
Agg            & $\rightarrow$ & \textsf{Sum}($\ts, a', b'$)   & $\sum_{i=a'}^{b'}\ts[i]$, where $[a',b']\subseteq[a,b]$\\
%               & |             & \textsf{Sum}($\timesOP, a, b$)   & $\sum_{i=a}^{b}(\tsi{1}[i]\times \tsi{2}[i])$\\
\hline
%\multicolumn{4}{l}{}\\
\hline
\multicolumn{4}{|l|}{\cellcolor{gray!15} \text{Time Series Expression (TSE)}}\\
\hline
$\bm{T}$              & $\rightarrow$& \text{input time series} & \\
%               & |           & \textsf{\timesOP}          & $(a, b, [d^{T_1}_{a}\times d^{T_2}_{a}, ...,d^{T_1}_{b}\times d^{T_2}_{b}])$\\
               & |          & \ser$(\upsilon, a, b)$    & $(a, b, [\underbrace{\upsilon,\upsilon,...,\upsilon}_{b-a+1}])$\\
               & |          & \shift$(\bm{T}, k)$             & $(a+k, b+k, [\bm{T}[a], ...,\bm{T}[b]])$\\
               & |           & $\plusOP$           & $(a, b, [\bm{T_1}[a]+\bm{T_2}[a], ...,\bm{T_1}[b]+\bm{T_2}[b]])$\\
               & |           & $\minusOP$          & $(a, b, [\bm{T_1}[a]-\bm{T_2}[a], ...,\bm{T_1}[b]-\bm{T_2}[b]])$\\
               & |           & $\timesOP$   & $(a, b, [\bm{T_1}[a]\times \bm{T_2}[a], ...,\bm{T_1}[b]\times \bm{T_2}[b]])$\\
\hline
\end{tabular}
\caption{Grammar of time series analytic (TSA). Let  $\tsi{1}= (a_1, b_1, [\bm{T_1}[a_1], ...,\bm{T_1}[b_1]])$ and $\tsi{2}= (a_2, b_2, [\bm{T_2}[a_2], ...,\bm{T_2}[b_2]])$ be the input time series in the time series expressions, $a=\max(a_1,a_2)$ and $b=\min(b_1, b_2)$.}
\label{table:query}
\end{table}
}\normalsize

\noindent\textbf{Time Series}
A time series  $\ts= (a, b, [\ts[a], \ts[a+1], ...,\ts[b]])$, $a\in N$, $b\in N$, is a sequence of data points $[\ts[a], \ts[a+1], ...,\ts[b]]$ observed from start time $a$ to end time $b$ ($a, b \in N$). Following the assumptions in~\cite{DBLP:conf/sigmod/MorseP07,DBLP:conf/vldb/ChenN04,DBLP:conf/icde/VlachosGK02} we assume that time is discrete and the resolution of any two time series is the same. Equivalently, we say $\ts$ is fully defined in the integer time domain $[a,b]$. We assume a domain $[1,n]$ is the global domain meaning that all the time series are defined within subsets of this domain.  When the domain of a time series $\ts$ is implied by the context, then $\ts$ can be simplified as $\ts= [\ts[a], \ts[a+1], ...,\ts[b]]$. 
%In addition, following the prior work~\cite{DBLP:conf/sigmod/MorseP07}, all the time series are normalized before being further processed.~\footnote{The standard normalization scheme can be found in~\cite{DBLP:conf/cp/GoldinK95}.}
\begin{example}
\label{example:time_series}
Assume the global domain is $[1,100]$. Consider two time series $\tsi{1} = (1,5,[61.52,59.54,58.64,59.36,$ $60.44])$ and $\tsi{2} = (3,6,[1.02, 1.03, 1.02, 1.02])$. Then  $\tsi{1}$ and $\tsi{2}$ are fully defined in domains [1,5] and [3,6] respectively. $\tsi{2}[4]=1.03$ refers to the $2^{nd}$ data point of $\tsi{2}$ at the $4$-th position in the global domain.
\end{example}

\rone{D3. Notation for restriction is confusing. Why is
$T|_{[a',b']}[i] = T[i]$? I thought i was the index of the data point.}

\rone{D4. 'For ease of exposition we omit the exact time points' -- I don't follow. do you assume equally spaced?}

\rone{D5. What does Shift($T_1,k$) mean? Do you assume k is a multiple of the resolution? Or that the resolution is 1?}

\begin{table*}[t]
\renewcommand{\tabcolsep}{0.2mm}
\centering
\footnotesize
\begin{tabular}{|B|c|c|E|} \hline
 TSA Expression                             & Definition                                                           & Equivalent TSA Expression                                              & Usage of error measures\\\hline
Average $\mu_{T_1}$  `$\mu$($T_1$)' & $\dfrac{1}{b_1-a_1+1}(\sum\limits_{i=a_1}^{b_1}T_1[i])$                             & $\dfrac{1}{b_1-a_1+1}$(Sum($\bm{T_1}$))                               & $\tes{T_1}$\\\hline
Standard Deviation $\sigma_{T_1}$ `$\sigma$($T_1$)'   & $\sqrt{\frac{1}{b_1-a_1+1}\Big(\sum\limits_{i=a_1}^{b_1}(T_1[i]-\mu_{T_1})^2\Big)}$ & $\sqrt{\frac{1}{b_1-a_1+1}\times Sum(\bm{T_1-}\ser(\mu_{T_1}))}$ & $\tes{T_1}$ \\\hline
Correlation $r_{(T_1,T_2)}$ `Corr($T_1$,$T_2$)'      & $\dfrac{\sum\limits_{i=max(a_1,a_2)}^{min(b_1,b_2)}\Big((T_1[i]-\mu_{T_1})(T_2[i]-\mu_{T_2})\Big)}{\sigma_{T_1}\times \sigma_{T_2}}$
																							               & $\dfrac{Sum\Big((\bm{T_1-}\ser(\mu_{T_1}))\bm{\times}(\bm{T_2-}\ser(\mu_{T_2}))\Big)}{\sigma_{T_1}\times \sigma_{T_2}}$ & $\fes{T_1}$,$\ses{T_1}$,$\tes{T_1}$, $\fes{T_2}$,$\ses{T_2}$,$\tes{T_2}$\\\hline
Cross-correlation $r_{(T_1,T_2,m)}$ `CCorr($T_1$,$T_2$,$m$)' & $\dfrac{\sum\limits_{i=max(a_1,a_2+m)}^{min(b_1,b_2+m)}\Big((T_1[i]-\mu_{T_1})(T_2[i+m]-\mu_{T_2})\Big)}{\sigma_{T_1}\times \sigma_{T_2}}$
                                                                                                             &$\dfrac{Sum\Big((\bm{T_1-}\ser(\mu_{T_1}))\bm{\times}(Shift(\bm{T_2},m)\bm{-}\ser(\mu_{T_2}))\Big)}{\sigma_{T_1}\times \sigma_{T_2}}$ & $\fes{T_1}$,$\ses{T_1}$,$\tes{T_1}$, $\fes{T_2}$,$\ses{T_2}$,$\tes{T_2}$\\\hline
Auto-correlation $r_{(T_1,m)}$  `ACorr($T_1$,$m$)'      & $\dfrac{\sum\limits_{i=a_1+m}^{b_1}\Big((T_1[i]-\mu_{T_1})(T_1[i+m]-\mu_{T_1})\Big)}{\sigma_{T_1}^2}$ 	
                                                                                                             &$\dfrac{Sum\Big((\bm{T_1-}\ser(\mu_{T_1}))\bm{\times}(Shift(\bm{T_1},m)\bm{-}\ser(\mu_{T_1}))\Big)}{\sigma_{T_1}\times \sigma_{T_1}}$ & $\fes{T_1}$,$\ses{T_1}$,$\tes{T_1}$\\\hline
\end{tabular}
\caption{Example TSA's for common statistics.  Let  $\tsi{1}= (a_1, b_1, [\cdots])$ and $\tsi{2}= (a_2, b_2, [\cdots])$ be the input time series in the time series analytic.}
\label{tbl:common-statistics-queries}
\end{table*}
\normalsize

% \begin{figure}[t]
% \center
% \includegraphics[width=0.4\textwidth]{Figure/query_language.png}
% \caption{Grammar of query expressions. Let $\bm{T}_1= (a, b, [d^{T_1}_{a}, ...,d^{T_1}_{b}])$ and $\bm{T}_2= (a, b, [d^{T_2}_{a}, ...,d^{T_2}_{b}])$ be the input time series in the time series expressions.}
% \label{fig:query}
% \end{figure}

\noindent\textbf{Time Series Analytic (TSA) Expressions} Table~\ref{table:query} shows the formal definition of the \textit{time series analytic} (called \textit{TSA}). The TSAs supported are expressions composed of linear algebra operators and arithmetic operators. Typically, the TSA has subexpressions that compose one or more linear algebra operators over multiple time series vectors as defined below.

%We support queries arithmetically and arbitrarily combining (i) the aggregations of the following time series operators and (ii) the inner product of two time series.

\begin{figure}[t]
\center
\includegraphics[width=0.45\textwidth]{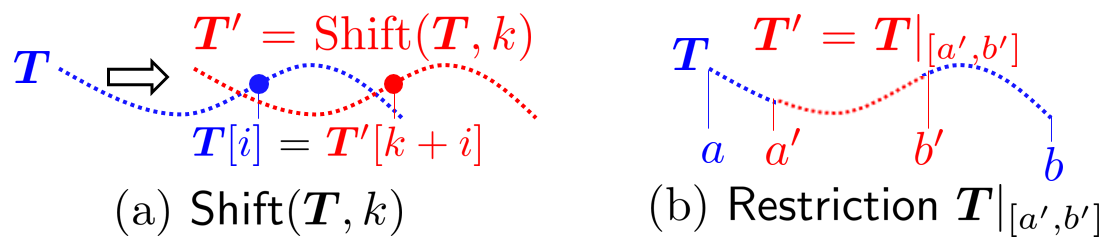}
\vspace{-4mm}
\caption{Time series \textit{Shift} and \textit{Restriction} operators.}
\label{fig:shift_restriction}
\end{figure}

\begin{compact_item}
  \item Given a numeric value $\upsilon$ and two integers $a$ and $b$, \ser$(\upsilon, a, b) = (a, b, [\upsilon,...,\upsilon])$. For example, \ser($1.6, 3,5$) produces $(3,5,[1.6,1.6,1.6])$.
  \item Given a time series $\ts=(a, b,  [\ts[a],...,\ts[b]])$ and an integer value $k$, \shift($\ts,k$)=$(a+k, b+k,[\ts[a], ...,\ts[b]])$. Notice \shift($\ts,k$)$[i+k]$ = $\ts[i]$ for all $a\leq i \leq b$. Figure~\ref{fig:shift_restriction}(a) visualizes the \shift{} operator. Consider the time series $\ts=(1,3,  [1.8, 1.6, 1.6])$, then \shift($\ts, 6$) is $(7, 9,$ $[1.8, 1.6, 1.6])$.
%  \item  Given two time series $\tsi{1} = (a, b, [\tsi{1}[a],...,\tsi{1}[b]])$ and $\tsi{2} = (a, b, [\tsi{2}[a],...,\tsi{2}[b]])$, $\plusOP = (a, b, [\tsi{1}[a]+\tsi{2}[a],$ $...,\tsi{1}[b]+\tsi{2}[b]])$. For example, consider $\tsi{1}= (1, 2, [3.3,3.5])$ and  $\tsi{2} = (1, 2, [1.0,1.2])$, then $\plusOP= (1,2, [4.3, 4.7])$.
%  \item  Given two time series $\tsi{1} = (a, b, [\tsi{1}[a],...,\tsi{1}[b]])$ and $\tsi{2} = (a, b, [\tsi{2}[a],...,\tsi{2}[b]])$,  $\minusOP = (a, b, [\tsi{1}[a]-\tsi{2}[a],$ $...,\tsi{1}[b]-\tsi{2}[b]])$.  For example, consider $\tsi{1} = (1, 2, [3.3,3.5])$ and  $\tsi{2} = (1, 2, [1.0,1.2])$ then $\minusOP = (1,2, [2.3, 2.3])$.
  \item  Given two time series $\tsi{1} = (a_1, b_1, [\tsi{1}[a_1],...,\tsi{1}[b_1]])$ and $\tsi{2} = (a_2, b_2, [\tsi{2}[a_2],...,\tsi{2}[b_2]])$,  $\timesOP = (a, b, [\tsi{1}[a]$ $\times\tsi{2}[a],$ $...,\tsi{1}[b]\times\tsi{2}[b]])$ where $a=max(a_1, a_2)$ and $b=min(b_1,b_2)$.~\footnote{Setting $a=max(a_1, a_2)$ and $b=min(b_1,b_2)$ ensures all the data points in $\timesOP$ are defined.}  For example, given $\tsi{1} = (1, 2, [3.3,$ $3.5])$ and  $\tsi{2} = (1, 2, [1.0,1.2])$ then $\timesOP = (1,2, [3.3, 4.2])$. Similarly, we define $\plusOP$ and $\minusOP$.
\end{compact_item}

\yannisp{TO YANNIS: COMBINE WITH EXAMPLE DISCUSSION.}
A time series analytic (TSA) is an arithmetic expression of the form $Arr_1 \otimes Arr_2 \otimes \ldots Arr_n$, where $\otimes$ are the standard arithmetic operators ($+, -, \times, \div, \sqrt{~}$) and $Arr_i$ is either an arithmetic literal or an aggregation  over a  time series expression. An aggregation expression \textsf{Sum}$(\bm{T}, a', b')$ computes the summation of the data points of $T$ in the domain $[a', b']$, i.e., \textsf{Sum}$(\ts, a', b')$=$\sum_{i=a'}^{b'}\ts[i]$ where $\ts$ can be an input time series or a derived time series computed by time series expressions (TSEs).~\footnote{Note that, when the time series expressions involve time shifting, we assume that the aggregation will only operator in the valid data points, that is the data points in the defined range.} 
When the bounds of $a'$ and $b'$ are implied from the context, we simplify  $Sum(\ts, a', b')$ to $Sum(\ts)$. 
%Note that the time series in \textsf{Sum}$(\ts, a, b)$ could either be a base time series or a derived time series that was computed from a set of base time series by applying time series operators.By combining and composing the basic operators, many common statistics queries over time series can be expressed. Table~\ref{tbl:common-statistics-queries} lists common statistics queries that can be expressed in this subset of linear algebra.

%For example, the cross-covariance $r_{(T_1,T_2,m)}$ is
%$\frac{1}{n}Sum((\bm{T_1-}Serialize(\mu_{T_1}))\bm{\times}(Shift(\bm{T_2},m)\bm{-}Serialize(\mu_{T_2})))$, where $\mu_{T_1}=(\sum_{i=1}^nT_1[i])/n$ and $\mu_{T_2}=(\sum_{i=1}^nT_2[i])/n$.
%More common statistics queries that can be expressed in this subset of linear algebra  are listed in Table~\ref{tbl:common-statistics-queries}.

\section{Internal, Compressed Time Series Representation}
\label{sec:time_series_index}
When a user inserts a time series into the database, \db{} physically stores the \textit{compressed time series representation} instead of the raw time series. More precisely, the user provides (i) a time series $T$, (ii) the identifier of a segmentation algorithm, which is chosen from a list provided by \db, and  (iii) the identifier of a function family, which is selected from a list provided by \db. Internally, \db{} uses the chosen segmentation algorithm and the chosen compression function family to partition  $T$ into a list of disjoint segments $T^1,...,T^n$. For each segment $T^i=(a, b, [T^i[a],...,T^i[b]])$, instead of storing its original data points $[T^i[a],...,T^i[b]]$, \db{} stores a \textit{compressed segment representation} $\rep{T^i}=(a, b, \rep{\ef{T}}, \Phi(T))$, where $a$ is the start position, $b$ is the end position, $\rep{\ef{T}}$ is the function representation of $\ef{T}$, where $\ef{T}$ is the estimation function chosen from the identified function family and $\Phi(T)$ is a set of (two to three depending on the function family) \textit{error measures}. 

Overall, for a time series $T$, \db{} physically stores (i) the list $L_T$=($\rep{T^1},...,\rep{T^n}$), and (ii) one token (which can simply be an integer) as the function family identifier.~\footnote{It is not necessary for \db{} to physically store a token for the segmentation algorithm identifier as the time series stored in \db{} has been partitioned already.}

We comment on the prior state-of-the-art segmentation / compression algorithms that \db{} uses in Appendix~\ref{appendix:segmentation_alg}. Next, we introduce the selection of the estimation function and the computation of error measures.
%in Section~\ref{subsec:estimation_function} and Section~\ref{subsec:error_measures} respectively.

\subsection{Estimation Function Selection}
\label{subsec:estimation_function}
\rone{D6. Section 3.1. Please define estimation function precisely.}
\rtwo{ When considering time series that have been compressed in advance (as explained in the first contribution of the paper), the error parameter should be computed before query evaluation. To compute the error, the (original) times series for each segment should be accessed and pre-processed. In practice, such pre-processing costs vs the cost of re-segmenting the time series with more appropriate compression functions / alignments should be discussed. }

Choosing an estimation function for a time series segment has two steps: (i) user identifies the function family, and (ii) \db{} selects the best function in the family, i.e., the function that minimizes the  Euclidean distance between the original values and the estimated values produced by the  function. 

\noindent\textbf{Step 1: Function family selection.}
%The user~\footnote{Usually the user is a domain expert.}  chooses a function family $\ff$ that is used for compressing a time series. In the present work, \db{} uses the same function family for all the segments of a time series. Thus, the time series physical representation stores a token $\tau$ that identifies the function family. 
Table~\ref{table:funciton_family_identifier}  gives example function family identifiers, which the user may select, and the corresponding function expressions. For example, $\tau$=``$p_2$'' means that the chosen function family is the ``second-degree polynomial function family'' and the corresponding function family expression is $\{ax^2+bx+c|a,b,c\in R\}$.

\noindent\textbf{Step 2: Estimation function selection.}
Any function $f$ in the chosen function family $\ff$ is a \textit{candidate estimation function}. 
%\db{} selects one \textit{estimation function} $\ef{T}$ among all the candidate estimation functions for a time series segment $T$.
Following the prior work~\cite{conf/icde/LazaridisM03,journals/is/AghabozorgiST15}, \db{} selects the candidate estimation function that minimizes the Euclidean distance between the original values and the estimated values produced by the  function to be the final estimation function. More precisely,
\begin{align}\label{equ:euclidean}
\ef{T} = \arg \min_{f \in \ff}\Big(\sum_{i=a}^{b}(\ts[i] - f(i))^2\Big)^{1/2}
\end{align}

% \begin{figure}[t]
% \center
% \includegraphics[width=0.35\textwidth]{Figure/estimation_function.png}
% \caption{Example of selecting estimation function.}
% \label{fig:estimation_function}
% \end{figure}

\begin{example}
\label{example:estimation_function}
Given a time series $\ts=(1,5,[0.2,0.4,0.4,$ $0.5,0.6])$, assume the function family identifier is ``$p_1$'' (i.e., ``first-degree polynomial function family''). Functions $f_1=0.05\times i+0.3$ and $f_2=0.09\times i+0.15$  are two candidate estimation functions. Finally, \db{} selects $f_2=0.09\times i+0.15$  as the estimation function since it produces the minimal Euclidean error, i.e., $0.0837$. 
\end{example}

\noindent\textbf{Function Representation (Physical) vs. Function (Logical)}.
Once an estimation function $\ef{T}$ is selected, \db{} stores the corresponding \textit{function representation} $\rep{\ef{T}}$, which includes 
%(i) the start and end positions of $T$,  
(i) the coefficients of the function $\ef{T}$, and 
(ii) the function family identifier $\tau$.~\footnote{All the segments in the same time series share one token $\tau$.} For example, the function representation of the estimation function in Example~\ref{example:estimation_function} is $\rep{\ef{T}}$ = ($(0.09, 0.15)$ , $p_1$) where $p_1$ is a function family identifier indicating that the function family is ``1-degree polynomial function family''.

When we talk about the function itself logically, it can be regarded as a vector that maps time series: given a domain $[a,b]$, the vector $[f(a), f(a+1), \ldots, f(b)]$ maps a value to each position in the domain $[a,b]$. For example, consider the estimation function $\ef{T}=0.09\times i+ 0.15$ in Example~\ref{example:estimation_function}. Then $\ts-\ef{T} = [0.2- \ef{T}(1), 0.4-\ef{T}(2), 0.4-\ef{T}(3), 0.5-\ef{T}(4), 0.6-\ef{T}(5)]= [0.2-0.24, 0.4-0.33,0.4-0.42,0.5-0.51,0.6-0.6]=[0.04,0.07,-0.02,-0.01,0]$.

\subsection{Error Measures}
\label{subsec:error_measures}
In addition to the estimation function, \db{}  stores extra \textit{error measures} $\es{T}=\{\fes{T}, \ses{T}, \tes{T}\}$ for each  time series segment $T$ (defined in domain $[a,b]$) where $\fes{T}$, $\ses{T}$, and $\tes{T}$ are defined in Table~\ref{tab:error_measures}.
\eat{
as follows:
\begin{compact_item}
\item $\fes{T} =\sqrt{\sum_{i=a}^b (T[i]-f^*_T(i))^2}$, which is  the $L_2$-norm of the estimation errors.
\item $\ses{T} =\sqrt{\sum_{i=a}^b (\ef{T}(i))^2}$, which is the $L_2$-norm of the estimated values produced by mapping the estimation function $\ef{T}$ to each position in the domain $[a,b]$.
\item $\tes{T} = \Big|\sum_{i=a}^b T[i]  - \sum_{i=a}^b f^*_T(i)\Big|$, which is called the \textit{reconstruction error}. It is the absolute difference between the sum of  original values and that of estimated values.
\end{compact_item}
}

\begin{example}
\label{example:error_measures}
Consider the time series $\ts=(1,5,[0.2,0.4,0.4,$ $0.5,0.6])$ in Example~\ref{example:estimation_function} again. $\ef{T}=0.09\times i+0.15$  is the estimation function. Thus $\fes{T}= \sqrt{\sum_{i=1}^5 (T[i]-f^*_T(i))^2}=0.0837$, 
$\ses{T}=\sqrt{\sum_{i=1}^5 (f^*_T(i))^2}=0.9813$, and 
$\tes{T}=|\sum_{i=1}^5 T[i]  - \sum_{i=1}^5 f^*_T(i)|=2.1-2.1=0$.
\end{example}

\noindent\textbf{Elimination of $\tes{T}$.} We will see in Lemma~\ref{lemma:orthogonal} (Section~\ref{sec:error_guarantee_align_VS}) that if the selected function family forms a vector space, then $\tes{T}$ is guaranteed to be $0$. Then we can avoid storing it.

\eat{
\begin{example}
\label{example:compressed_segment_representation}
Consider the time series $T_1$ in  Figure~\ref{fig:intro_time} again. The compressed segment list index $L_{T_1}= \{\rep{T_1^1}, \rep{T_1^2}\}$ where  $\rep{T_1^1}$ and $\rep{T_1^2}$ are the compressed segment representations of $T_1^1$ and $T_1^2$ respectively. More precisely,  (i) $\rep{T_1^1}=(1, 112,\rep{\ef{T_1^1}}, \Phi(T_1^1))$ where $\rep{\ef{T_1^1}}=((-0.072, 69.38),p_1)$  and $\Phi(T_1^1)$ = $(0.55, 411.82, 0)$; and (ii) $\rep{T_1^2}=(113, 600, \rep{\ef{T_1^2}}, \Phi(T_1^2))$ where $\rep{\ef{T_1^2}}=((-0.002, 65.77),p_1)$  and $\Phi(T_1^2)$ = $(1.77, 827.31, 0)$. 
\end{example}
}

% \begin{example}
% Consider the temperature time series $T$ in Figure~\ref{fig:intro_time}. \db{} partitions $T$ into two segments $T^1$ and $T^2$. For $T^1$ (resp. $T^2$), the estimation function is $f^*_{T^1} = -0.072i+69.38$ (resp. $f^*_{T^2} = -0.002i+65.77$) and the corresponding set of error measures is $(0, 1.05, 844.5)$ (resp. $(0, 1.77, 1338.2)$). Therefore, \db{} finally stores $\tindex=\{$((0.072, 69.3), (0, 1.05, 844.5)),((-0.002, 65.77), (0, 1.77, 1338.2))$\}$ for $T$. 
% \end{example}
\section{Error guarantee computation}
\label{sec:error_estimation}

\rone{D1. Paper is full of formalism and hard to follow, due to lack fo examples.}

\rone{D8. Section 4, the error guarantees all seem to be cauchy-schwarz type inequalities. If so, please consider explaining the theorems this way, because I found the current presentation hard to follow (eg 4.1.1).}

\rtwo{The results on AI compliance (Section 4) are nicely introduced, although not surprising. }

\rtwo{The conclusion that may be drawn from Section 4 could be to consider appropriate compression schemes instead of trying to compute errors for compressions schemes which are not in AI. It is not clear why compression functions which would not comply with the AI requirement (typically, other than LSF) are considered. And if LSF is considered, we expect discussions related to the structures/indexes and query processing strategies which should be used to integrate the error parameters at query time. }

%\begin{definition}[Tight error guarantee]
%Given a query $q$, let $\mathds{E}$ be the collection of error measures of the segments in the segmentations of the time series $T_1,...,T_n$ involved by $q$, then we say that \textit{$\hat{\varepsilon}$ is a tight error guarantee of $q$ that uses $\mathds{E}$} if there exists an instance segmentations of $T_1,...,T_n$ such that the true error $\varepsilon$ equals to $\hat{\varepsilon}$.
%\end{definition}

\noindent\textbf{Error Guarantee Definition.}
Given a TSA $q$ involving time series $T_1,..,T_n$, let $R$ be the accurate answer of $q$ by executing $q$ directly on the original data points of $T_1,..,T_n$. Let  $\hat{R}$ be the approximate answer of $q$ by executing $q$ on the compressed time series representations. Then $\te=|\hat{R}-R|$ is the \textit{true error} of $q$. Notice that $\te$ is unknown since $R$ is unknown. An upper bound $\hat{\varepsilon}$ ($\hat{\varepsilon}\geq \varepsilon$) of the true error is called a \textit{deterministic error guarantee} of $q$. With the help of $\eg{}$, we know that the accurate answer $R$ is within the range $[\hat{R}-\hat{\varepsilon}, \hat{R}+\hat{\varepsilon}]$ with \textit{$100\%$} confidence. \db{} provides \textit{tight} deterministic error guarantees for time series expressions defined in Table~\ref{table:query} (Section~\ref{sec:DataQueries}).

\noindent\textbf{Error Guarantee Decomposition.}
Recall that the time series analytic $q$ defined in Table~\ref{table:query} (Section~\ref{sec:DataQueries}) combines one or more time series aggregation operations via arithmetic operators, i.e., $q = Agg_1\otimes Agg_2\otimes\cdots \otimes Agg_n$ where $\otimes\in\{+,-,\times,\div, \sqrt{~}\}$. In order to provide the deterministic error guarantee $\eg{}$ of the time series analytic $q$, the key step is to calculate the deterministic error guarantee $\eg{Agg_i}$ of each aggregation operation $Agg_i$.  Once we have  $\eg{Agg_i}$ for each aggregate expression, it is not hard to combine them to get the final error guarantee (see  Appendix~\ref{appendix:propagate_errors}). 

Given a TSA $Agg= \textsf{Sum}(\bm{T})$ and the compressed time series representation $L_T = \{\rep{T^1},...,\rep{T^k}\}$. When calculating $\eg{Agg}$, there are two cases depending on whether $\ts$ is an input time series or not.~\footnote{If a time series is generated by applying some time series operators, then it is not a base time series. For example, $\ts = \timesOP$, then $\ts$ is not a base time series.}

\begin{figure}[t]
\center
\includegraphics[width=0.5\textwidth]{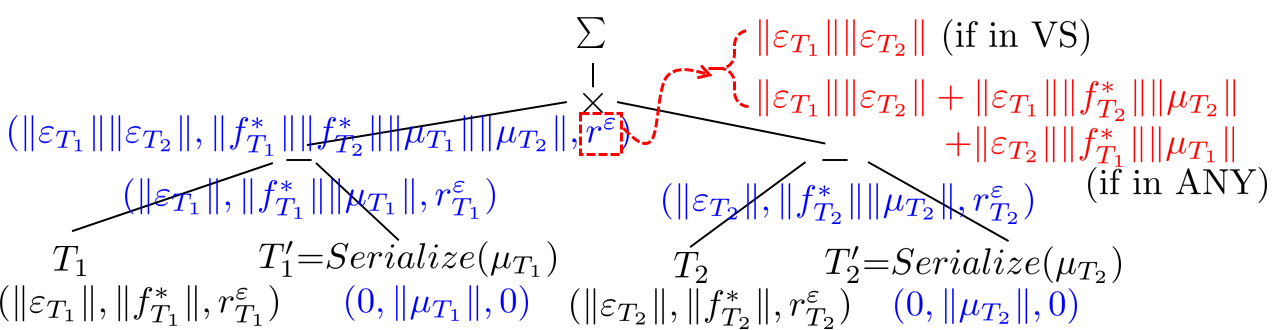}
\caption{Example of error measures propagation. Error measures in black color are precomputed offline during insertion time, while error measures in blue color are computed during the TSA processing time. The final error guarantees are in red color.}
\label{fig:error_measures_propogation}
\center
\includegraphics[width=0.5\textwidth]{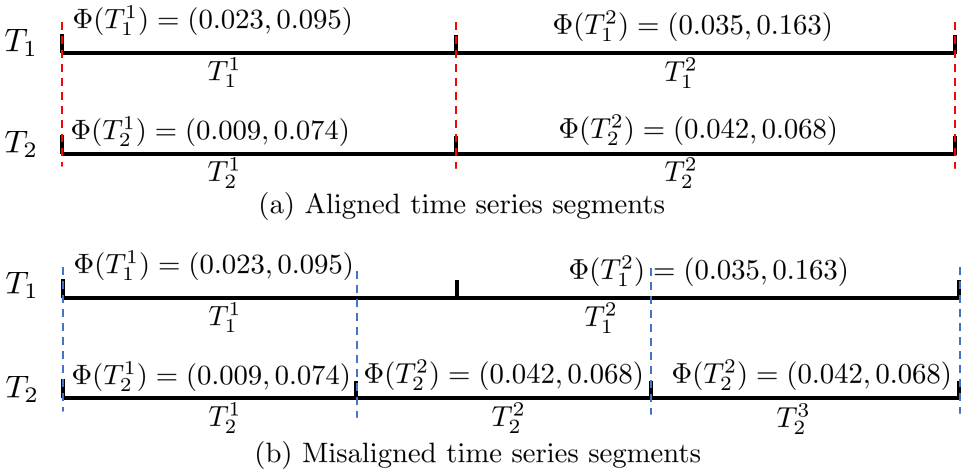}
\caption{Example of aligned segments and misaligned segments.}
\label{fig:aligned_misaligned_example}
\end{figure}

\begin{compact_item}
\item Case 1. $\ts$ is an input time series, then $\eg{Agg} = \sum_{i=1}^k \tes{T^i}$ where $\tes{T^i}$ is the reconstruction error in the error measures of $T^i$.~\footnote{Here we assume the aggregation operator aggregates the whole time series.}
\item Case 2. $\ts$ is a derived time series by applying the  time series operators (recursively), \textsf{\ser($\upsilon,a,b$)}, \textsf{Shift($\ts, k$)}, \textsf{$\plusOP$}, \textsf{$\minusOP$} and \textsf{$\timesOP$}. In this case, the aggregation operator $Agg = Sum(\ts)$ can be depicted as a tree. Figure~\ref{fig:error_measures_propogation} shows an example tree of the aggregation operator in the ``correlation TSA''. In order to  compute $\eg{Agg}$, we first calculate the error measures $\es{\ts}=(\fes{T}, \ses{T}, \tes{T})$ for the root time series in the tree by propagating the error measures from the bottom time series to the root. Then we return the $\tes{T}$ in the $\es{\ts}$ as the final error guarantee.  
\end{compact_item}

Next, we focus on computing the error measures for derived time series. 
We first explain the simpler case where each time series is a single segment. Table~\ref{tbl:error-guarantee-propagation} shows the formulas for computing error measures for derived time series in this case. For the general scenario where multiple segments are involved in each input time series in the expression, there are two cases depending on whether the segments are aligned or not: If the $i$-th segment in $T_1$ has the same domain with the $i$-th segment in $T_2$ for all $i$, then $T_1$ and $T_2$ are \textit{aligned}, otherwise, they are \textit{misaligned}. 

\yannisp{TO YANNIS: WHY TOGETHER SUM AND $T_1\times T_2$}
In the following, we will show how to compute the most challenging error guarantee $\eg{Sum(T_1\times T_2)}$ in both aligned and misaligned cases in Section~\ref{sec:error_guarantee_aligned} and Section~\ref{sec:error_guarantee_misaligned} respectively. The computation of error guarantees of other expressions (i.e., \textsf{\ser($\upsilon,a,b$)}, \textsf{Shift($\ts, k$)}, \textsf{$\plusOP$} and \textsf{$\minusOP$}) is presented in Appendix~\ref{sec:error_guarantee_others}.

\subsection{Error Guarantee on Aligned Segments}
\label{sec:error_guarantee_aligned}
\noindent\textbf{Notations.}
Given a time series $\ts=(\ts[a],...,\ts[b])$ and the estimation function $\ef{T}$ of $T$, $\bm{\varepsilon}_T = \ts-\ef{T} = (\ts[a]-\ef{T}(a),...,\ts[b]-\ef{T}(b))$ is the \textit{vector of errors} produced by the estimation function. In the following, $\ts$, $\ef{T}$ and $\bm{\varepsilon}$ are all regarded as vectors. $\langle f_1, f_2 \rangle = \sum_{i=a}^{b}f_1(i)f_2(i)$ is the inner product of $f_1$ and $f_2$. $V|_{[a,b]}$ is a \textit{restriction} operation, which restricts a vector $V$ to the domain [a,b]. Recall a time series segment is a subsequence of a time series. Thus, a segment is the \textit{restriction} of a time series $\ts$ from a bigger domain $[a,b]$ into a smaller domain $[a',b']\subseteq[a,b]$, denoted as $\ts|_{[a',b']}$. Figure~\ref{fig:shift_restriction}(b) visualizes the restriction operator. For example, consider a time series $\ts= (1,4,[1.2, 1.3, 1.3, 1.2])$, then $\ts|_{[2,3]}=(2,3,[1.3, 1.3])$ is a restriction of $\ts$. Note that $\ts|_{[a',b']}[i] = T[i]$ for all $i\in[a',b']$.  

Given two compressed time series representation $L_{T_1}=(\rep{T_1^1},...,\rep{T_1^k})$ and $L_{T_1}=(\rep{T_2^1},...,\rep{T_2^k})$ for the \textit{aligned} time series $T_1=(T_1^1,...,T_1^k)$ and $T_2=(T_2^i,...,T_2^k)$ where $T_1^i = T_1|_{[a_i,b_i]}$ and $T_2^i = T_2|_{[a_i,b_i]}$. Notice $T_1^i$ and $T_2^i$ have the same domain, i.e., $[a_i,b_i]$, for all $i\in [1,k]$. 
%Recall that \db{} stores the error measures $(\fes{T_i^j}, \ses{T_i^j}, \tes{T_i^j})$ for each segment $T_i^j$. 
%Connecting back to the restriction operation, we have  $\fes{T_1^i}=\|\bm{\varepsilon}_{T_1}|_{[a_i,b_i]}\|_2$ and $\|\ef{T_1^i}\|_2 = \|\ef{T_1}|_{[a_i,b_i]}\|_2$. 
For any estimation  function family, the error guarantee of \textsf{Sum}($\bm{T_1\times T_2}$) on aligned time series is: 
\begin{align}
\label{equ:aligned_ANY}
\varepsilon \leq \sum_{i=1}^{k}\Big(\fes{T_1^i}\fes{T_2^i}+\fes{T_1^i}\ses{T_2^i} + \ses{T_1^i}\fes{T_2^i}\Big)
\end{align}
The details are shown in Appendix~\ref{appendix:aligned_ANY}.

\begin{example} \label{example:aligned_any}
Consider the two aligned time series in Figure~\ref{fig:aligned_misaligned_example}(a). Both $T_1$ and $T_2$ are partitioned into two segments in this case, i.e., ($T_1^1, T_1^2$) and ($T_2^1, T_2^2$). \db{} stores the error measures $\es{T_i^j}$ for each segment $T_i^j$. For instance, $\es{T_1^1}=(\fes{T_1^1}, \ses{T_1^1}, \tes{T_1^1})=(0.023, 0.95, 0)$. Then the error guarantee of \textsf{Sum}($\bm{T_1\times T_2}$) on $T_1$ and $T_2$ is computed as $(\fes{T_1^1}\fes{T_2^1}+\fes{T_1^1}\ses{T_2^1} + \ses{T_1^1}\fes{T_2^1}) + (\fes{T_1^2}\fes{T_2^2}+\fes{T_1^2}\ses{T_2^2} + \ses{T_1^2}\fes{T_2^2})=(0.023\times 0.009+0.023\times 0.074 + 0.095\times 0.009)+(0.035\times 0.042+0.035\times 0.068 + 0.163\times 0.042)=0.01346$. 
%The true error in this case is  $0.001520$.
\end{example}

\subsubsection{Orthogonal projection optimization}
\label{sec:error_guarantee_align_VS}
If the estimation function family forms a vector space (\textsf{VS}),~\footnote{A vector space is a set that is closed under finite vector addition and scalar multiplication. \url{http://mathworld.wolfram.com/VectorSpace.html}.} then we can apply the \textit{orthogonal projection property} in VS to significantly reduce the error guarantee of  $sum(T_1\times T_2)$ from  Formula~\ref{equ:aligned_ANY} to Formula ~\ref{equ:aligned_VS}.
\begin{align}
\label{equ:aligned_VS}
\varepsilon
& = \Big|\sum_{i=1}^{k}\Big(\underbrace{\langle \bm{\varepsilon}_{T_1^i}, \ef{T_2^i} \rangle}_{=0~in~VS} +  \underbrace{\langle \bm{\varepsilon}_{T_2^i}, \ef{T_1^i} \rangle}_{=0~in~VS} + \langle \bm{\varepsilon}_{T_1^i}, \bm{\varepsilon}_{T_2^i} \rangle \Big) \Big|\nonumber\\
&\leq \sum_{i=1}^{k}\Big(\fes{T_1^i}\fes{T_2^i}\Big)
\end{align}
\begin{example}\label{example:aligned_vs}
Consider the two aligned time series in Figure~\ref{fig:aligned_misaligned_example}(a) again. The estimation function family is polynomial function family, it is  \textsf{VS}. Based on Formula~3, the error guarantee for \textsf{Sum}($\bm{T_1\times T_2}$) is $\fes{T_1^1}\times \fes{T_2^1}$ + $\fes{T_1^2}\times\fes{T_2^2} $= $0.023\times 0.009 + 0.035 \times 0.042 = 0.001677$. This error guarantee is about \textbf{$8\times$} smaller than that in Example~\ref{example:aligned_any} (i.e., $0.01346$), where we did not take into account that the function family is VS. 
\end{example}

\begin{figure}[t]
\center
\includegraphics[width=0.5\textwidth]{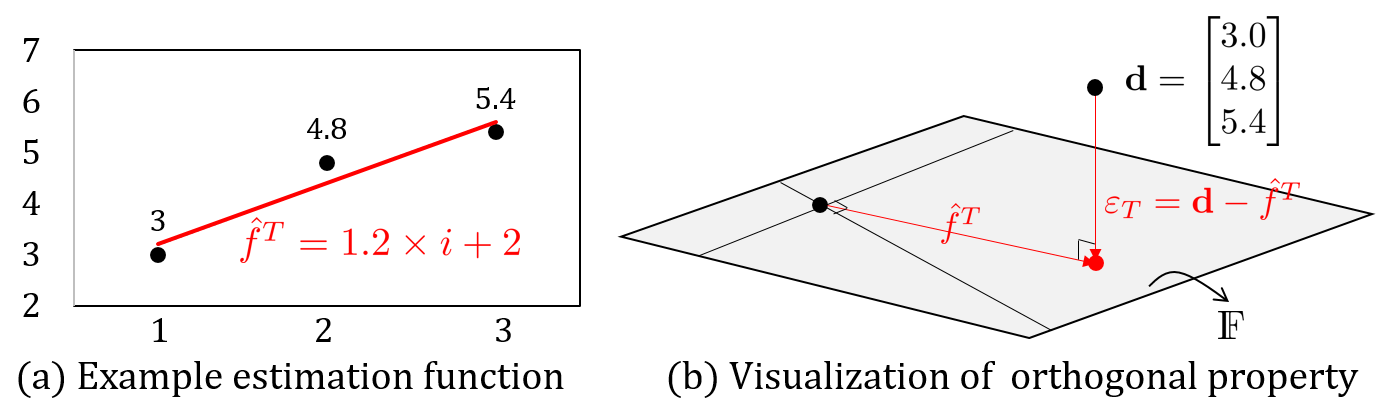}
\vspace{-2mm}
\caption{(a) shows the estimation function for three data points. (b) visualizes the  orthogonal projection of the three data points onto the 2-dimensional plane $\mathds{F}$.}
\label{fig:orthogonal_projection}
\center
\includegraphics[width=0.5\textwidth]{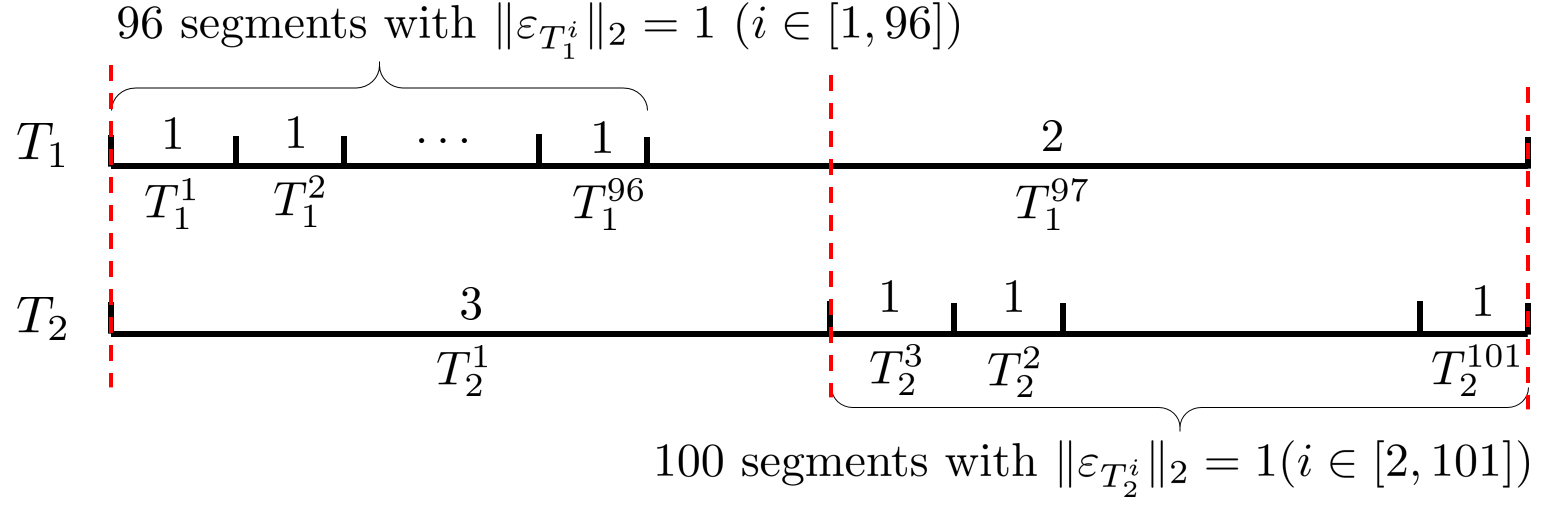}
\caption{Example of segment combination selection.}
\label{fig:segment_combination}
\end{figure}

\noindent\textbf{Orthogonal projection property}.
Example~\ref{example:aligned_vs} indicates the power of the orthogonal projection optimization. Lemma~\ref{lemma:orthogonal} 
is a proof of Formula~3.

\begin{lemma}
\label{lemma:orthogonal} (Orthogonal Projection Property)
Let $\mathds{F}$ be a function family forms a vector space \textsf{VS} and
$f^*_T\in \mathds{F}$ be the estimation function of time series $T$. Then $f^*_T$ is the orthogonal projection of $\mathbf{T}$ onto $\mathds{F}$~\cite{nelson1973probability}.
\end{lemma}

Lemma~\ref{lemma:orthogonal} implies that $\bm{\varepsilon}_T$ is orthogonal to any function $f_T\in\mathds{F}$, \yannisp{it was $f^T \in \mathds{F}$ but I don't think you really meant the unique $f^T$. Chunbin: Yes, it should be $f$.} which means  $\langle \bm{\varepsilon_T}, \bm{f_T}\rangle = 0$.
Therefore, given any two aligned segments  $T_1^i$ and $T_2^i$,
as both $f_{T_1^i}^*$ and $f_{T_2^i}^*$ are in VS, thus $\langle \bm{\varepsilon}_{T_1^i}, \ef{T_2^i} \rangle= 0$ and $\langle \bm{\varepsilon}_{T_2^i}, \ef{T_1^i} \rangle=0$.

%The example in Figure~\ref{fig:orthogonal_projection} illustrates the orthogonal property in VS and its application.
For visualization purposes, consider a time series with three data points  $T=(1,3,[3.0,4.8,5.4])$ and let $\mathds{F}$ be the 1-degree polynomial function family (i.e., 2-dimensional). The estimation function that minimizes the error to the original data is $\ef{T} = 1.2\times i+2$ (Figure~\ref{fig:orthogonal_projection}(a)).  As shown in Figure~\ref{fig:orthogonal_projection}(b), $\ef{T}$ is the orthogonal projection of $\bm{T}$ onto $\mathds{F}$. The error vector is $\bm{\varepsilon}_T = (-0.2, 0.4,-0.2)$. Based on Lemma~\ref{lemma:orthogonal}, for any candidate estimation function $f=\alpha\times i+ \beta$ $(\alpha, \beta\in R)$, we have
$\langle \bm{\varepsilon_T}, \bm{f}\rangle = 0.8\alpha - 0.8\alpha + 0.4\beta - 0.4\beta =0$.

\noindent\textbf{Elimination of $\tes{T}$.} We can get an extra benefit from the orthogonal projection property in saving space, i.e., the error measure $\tes{T}$ can be avoided as it is guaranteed to be $0$.
% \begin{lemma}
% \label{lemma:reconstruction_error_avoid}
% Let $\ff$ be a function family forms a vector space \textsf{VS} and
% $\ef{T} \in \ff$ be the estimation function of time series $T$. Then $\tes{T} = |\sum_{i=a}^b T[i] - \sum_{i=a}^b f^*_T(i)| = 0$ always hold.
% \end{lemma}
This is because $\tes{T} = \langle T-\ef{T}, 1\rangle$ and $1$ is a constant function in the function family in VS. According to  Lemma~\ref{lemma:orthogonal}, we know  $\langle T-\ef{T}, 1\rangle = 0$. Therefore, we have $\tes{T} = 0$.

\noindent\textbf{Amplitude-independent (AI)}. The orthogonal projection optimization can significantly reduce the error guarantees. It allows the error guarantees to get rid of the \textit{amplitudes} of the original time series values (referring to $\ses{T}$) by only consider the reconstruction error (referring to $\fes{T}$) of each time series. The error guarantees provided by \db{} in \textsf{VS} are called  \textit{amplitude-independent (AI)} error guarantees.

%In addition, in aligned environment, the  error guarantees in \textsf{VS} also applies to that in \textsf{LSF}, since \textsf{LSF} is a subset of \textsf{VS} (Section~\ref{subsec:estimation_function}). 

\subsection{Error Guarantee on Misaligned Segments}
\label{sec:error_guarantee_misaligned}

\rone{W1. Presentation is dense and hard to follow.
Misaligned time series i think is the main nontrivial aspect, 
and I could not follow it at all. Eg, Algorithm 1 is never explained.}

\eat{In reality, the time series involved in queries are usually misaligned due to the following two reasons:
First, variable length segments lead to better compression. Second, most useful queries operate on \textsf{Shift} time series. For example, the cross-correlation and the auto-correlation require shifting the time series. Even when the original time series segmentations are aligned, they are misaligned after shifting.}

Given two compressed time series representation $L_{T_1}=(\rep{T_1^1},...,\rep{T_1^{k_1}})$ and $L_{T_1}=(\rep{T_2^1},...,\rep{T_2^{k_2}})$ for the  misaligned time series $T_1=(T_1^1, ..., T_1^{k_1})$ and $T_2=(T_2^1, ..., T_2^{k_2})$ where the domains of $T_1^i$ and $T_2^i$ are $[a_i^1,b_i^1]$ and $[a_i^2,b_i^2]$ respectively. The major challenge in the misaligned case is that for a domain $[a_1^i,b_1^i]$,  the error measures of the segment $T_1|_{[a_1^i,b_1^i]}$ are precomputed, however, the error measures of the segment $T_2|_{[a_1^i,b_1^i]}$ may be unknown as $T_2|_{[a_1^i,b_1^i]}$ in general is not one of the segments $T_2^1,...,T_2^{k_2}$.

Let $\cover{T}{[a,b]}$ be the set of segments in $T$ covering the domain $[a,b]$. For example, consider the two misaligned time series $T_1$ and $T_2$ in Figure~\ref{fig:aligned_misaligned_example}(b), $\cover{T_2}{[a_1^1,b_1^1]}$ = $\{T_2^1,T_2^2\}$ as the segments $T_2^1$ and $T_2^2$ in $T_2$ cover the domain $[a_1^1,b_1^1]$.~\footnote{If time series $T_1$ and $T_2$ are aligned, then $\cover{T_2}{[a_1^i,b_1^i]}$ always returns one single segment.} If any kinds of function families are allowed, i.e., in ANY, the error guarantee $\hat{\varepsilon}$ of \textsf{Sum}($\bm{T_1\times T_2}$) on misaligned time series is:

\begin{figure}[h]
\center
\includegraphics[width=0.5\textwidth]{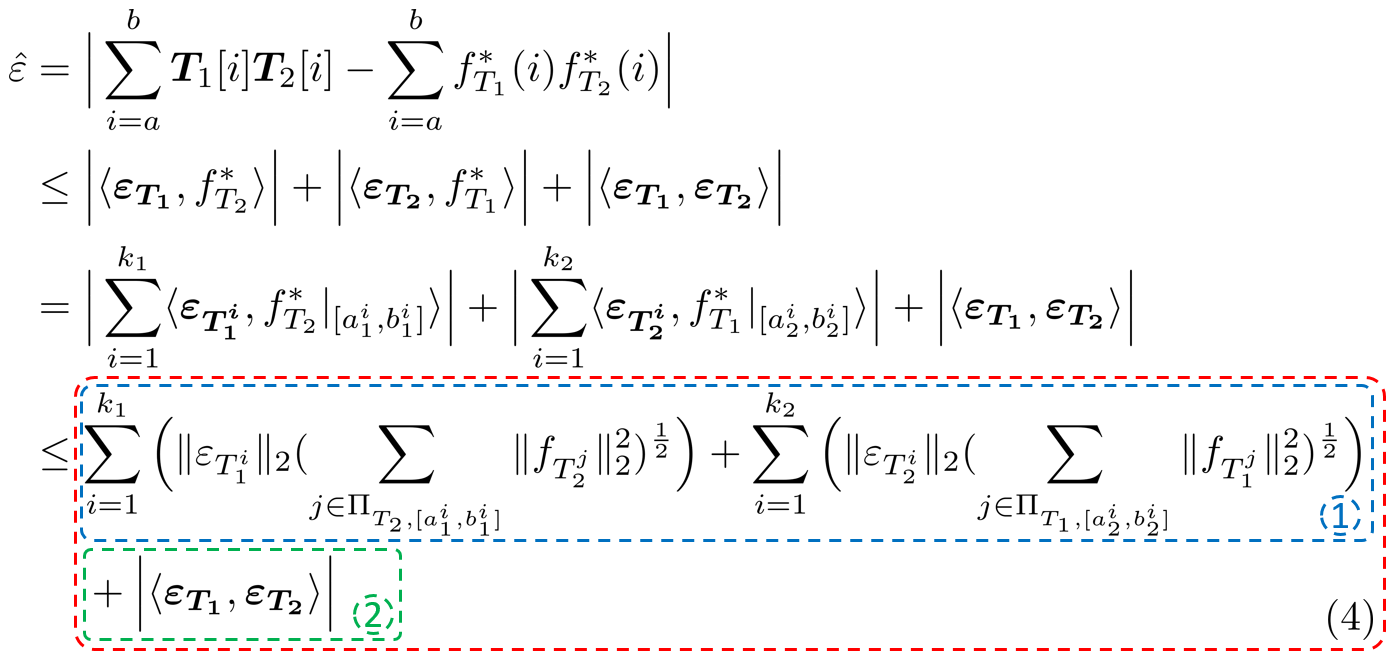}
\end{figure}

Formula~4 is a stepping stone towards producing the final formula as the computation of $|\langle \bm{\varepsilon_{T_1}}, \bm{\varepsilon_{T_2}}\rangle|$ (Formula~4$\textcircled{2}$) has not been given yet. It will be discussed in Section~\ref{sec:segment_combination_selection}. Section~\ref{sec:error_guarantee_misalign_LSF} discusses how to apply the orthogonal property optimization to improve Formula~4$\textcircled{1}$.

\subsubsection{Segment combination selection}
\label{sec:segment_combination_selection}
%If $T_1$ and $T_2$ are single segments, then $|\langle \bm{\varepsilon_{T_1}}, \bm{\varepsilon_{T_2}}\rangle|$ can be directly computed as $|\langle \bm{\varepsilon_{T_1}}, \bm{\varepsilon_{T_2}}\rangle|\leq \fes{T_1}\fes{T_2}$. However, both $T_1$ and $T_2$ have more than one segments. If $T_1$ and $T_2$ are aligned, then $|\langle \bm{\varepsilon_{T_1}}, \bm{\varepsilon_{T_2}}\rangle|$ can be calculated as $|\langle \bm{\varepsilon_{T_1}}, \bm{\varepsilon_{T_2}}\rangle| = \sum_{i=1}^k|\langle \bm{\varepsilon_{T_1|_{[a_1^i,b_1^i]}}}, \bm{\varepsilon_{T_2|_{[a_1^i, b_1^i]}}}\rangle|$ where $[a_1^i, b_1^i]$ and $[a_2^i, b_2^i]$ are domains of segments $T_1^i$ and $T_2^i$. If $T_1$ and $T_2$ are aligned, then $a_1^i = a_2^i$ and $b_1^i=b_2^i$. So $|\langle \bm{\varepsilon_{T_1|_{[a_1^i,b_1^i]}}}, \bm{\varepsilon_{T_2|_{[a_2^i, b_2^i]}}}\rangle|= |\langle \bm{\varepsilon_{T_1^i}}, \bm{\varepsilon_{T_2^i}}\rangle|\leq \fes{T_1^i}\fes{T_2^i}$. 
\yannisp{TO YANNIS: NEED TO THINK A BETTER SENTENCE FOR THIS}
To compute $|\langle \bm{\varepsilon_{T_1}}, \bm{\varepsilon_{T_2}}\rangle|$, one straightforward method (called IS) is to use the domains of segments in $T_1$ and $T_2$ independently, then choose the one with minimal value. Let's first see how to compute $|\langle \bm{\varepsilon_{T_1}}, \bm{\varepsilon_{T_2}}\rangle|$ with the domains of segments in $T_1$.
\begin{align*}
|\langle \bm{\varepsilon_{T_1}}, \bm{\varepsilon_{T_2}}\rangle|&
\leq \sum_{i=1}^{k_1}\big|\langle \bm{\varepsilon_{T_1|_{[a_1^i,b_1^i]}}}, \bm{\varepsilon_{T_2|_{[a_1^i,b_1^i]}}}\rangle\big|=\sum_{i=1}^{k_1}\big|\langle \bm{\varepsilon_{T_1^i}}, \bm{\varepsilon_{T_2|_{[a_1^i,b_1^i]}}}\rangle\big|\\
%&\leq \sum_{i=1}^{k_1}\big(\fes{T_1^i}\|\varepsilon_{T_2|_{[a_1^i,b_1^i]}}\|_2\big)\\
&\leq \sum_{i=1}^{k_1}\big(\fes{T_1^i}(\sum_{j\in\cover{T_2}{[a_1^i,b_1^i]}}\fes{T_2^j}^2)^{\frac{1}{2}}\big)\\
\end{align*}

In the last step of the above Formula, $T_2|_{[a_1^i, b_1^i]}$ is not a segment that \db{} precomputed in $T_2$. Thus, we need to use all the segments in $T_2$ covering $[a_1^i, b_1^i]$, i.e., $\cover{T_2}{[a_1^i, b_1^i]}$. Similarly, we can compute  $|\langle \bm{\varepsilon_{T_1}}, \bm{\varepsilon_{T_2}}\rangle|$ according to the domains of segments in $T_2$. Finally, IS chooses the minimal one between them. 
\eat{
That is, the output of IS is:
\begin{align*}
\min\Big(\sum_{i=1}^{k_1}\big(\fes{T_1^i}(\sum_{j\in\cover{T_2}{[a_1^i,b_1^i]}}&\fes{T_2^j}^2)^{\frac{1}{2}}\big), \\
&\sum_{i=1}^{k_2}\big(\fes{T_2^i}(\sum_{j\in\cover{T_1}{[a_2^i,b_2^i]}}\fes{T_1^j}^2)^{\frac{1}{2}}\big)\Big)
\end{align*}
}
However, IS does not produce tight guarantees, \db{} does not use it.  Next, we show the tight computation called \textit{OS}, which is used by \db.

\noindent\textbf{Optimal strategy (OS)}
OS (Algorithm~\ref{alg:optimal_segmentation})  first computes an error distribution array $E_{T_1}$ (resp. $E_{T_2}$) for $T_1$ (resp. $T_2$) (line 2) according to the domains of the segments as follows:
\begin{align}
E_{T_1} = \Big\{\fes{T_1^i}\times \Big(\sum_{j\in \cover{T_2}{[a_1^i,b_1^i]}}\fes{T_2^i}^2\Big)^{\frac{1}{2}}\Big|1\leq i \leq k_1\Big\}\nonumber\\
E_{T_2} = \Big\{\fes{T_2^i}\times \Big(\sum_{j\in \cover{T_1}{[a_2^i,b_2^i]}}\fes{T_1^i}^2\Big)^{\frac{1}{2}}\Big|1\leq i \leq k_2\Big\}\nonumber
\end{align}

\yannisp{FOR THE ``CURRENT''. WE NEED TO FIX THIS BUT NOT SURE HOW TO.THERE IS NO ``CURRENT'' IN THE ALGORITHM. I WONDER IF WE SHOULD ADD A VARIABLE CURRENT OR COMMENTS IN THE ALGORITHM THAT SHOW WHEN THE CURRENT CHANGES. GIVE IT A TRY.}
Then OS increases $\varepsilon_1$ (resp. $\varepsilon_2$) by adding the values from $E_{T_1}$  (resp.  $E_{T_2}$) (lines 4-7) and checks whether the current domain achieves the minimal errors (lines 8-17). If yes, OS adds the current domain (either $[start, b_1^{i_1}]$ or $[start, b_2^{i_2}]$) to the final segment combination list. After that, OS starts from a new domain and repeats the previous steps until all the segments are processed. The time complexity of OS is $O(k_1+k_2)$.

Let $OPT(L_{T_1}, L_{T_2})$ be the segment combination returned by OS. Then $|\langle \bm{\varepsilon_{T_1}}, \bm{\varepsilon_{T_2}}\rangle|$ is computed as follows:
\begin{align}
\label{equ:error_inner_product}
&|\langle \bm{\varepsilon_{T_1}}, \bm{\varepsilon_{T_2}}\rangle| \leq
\sum_{[a,b] \in OPT(L_{T_1}, L_{T_2})} \Big|\langle \bm{\varepsilon_{T_1}}|_{[a, b]}, \bm{\varepsilon_{T_2}}|_{[a, b]}\rangle \Big|\nonumber\\
 &\leq \sum_{[a,b] \in OPT(L_{T_1}, L_{T_2})} \Big(\Big(\sum_{i\in \cover{T_1}{[a,b]}} \|\bm{\varepsilon_{T_1^i}}\|_2^2\Big)^{\frac{1}{2}}\Big(\sum_{i\in \cover{T_2}{[a,b]}} \|\bm{\varepsilon_{T_2^i}}\|_2^2\Big)^{\frac{1}{2}}\Big)\nonumber
\end{align}

OS provides the  optimal segment combination that produces the minimum $|\langle \bm{\varepsilon_{T_1}}, \bm{\varepsilon_{T_2}}\rangle|$. The tightness proof is presented in Appendix~\ref{appendix:proof_optimal_segment_combination}.

\begin{example}
Consider the two misaligned time series in Figure~\ref{fig:segment_combination}. The value of $\fes{T_i^j}$ for each segment $T_i^j$ is labeled there. OS produces the segment combination $S=\{[a_1^1, b_2^1],[b_2^1, b_2^{101}]\}$ as visualized by the red lines. Then $|\langle \bm{\varepsilon_{T_1}}, \bm{\varepsilon_{T_2}}\rangle|= (3\times (96\times 1^2+2^2)^{\frac{1}{2}})+(2\times (100\times 1^2)^{\frac{1}{2}}) = 3\times 10+2\times 10 = 50$. 
However, IS outputs  $|\langle \bm{\varepsilon_{T_1}}, \bm{\varepsilon_{T_2}}\rangle| = min((3\times 96+2\times\sqrt{100+9}), (3\times\sqrt{96+2^2} + 100\times 2)) =\min(308.88, 230) = 230$, which is $4.6\times$ larger than the result returned by OS.
\end{example}
\begin{algorithm}[t]\small
\KwIn{Compressed segment representations $L_{T_1}$, $L_{T_2}$}
\KwOut{A segment combination $OPT$}
$\varepsilon_1= 0$, $\varepsilon_2=0$, $i_i= 0$, $i_2= 0$, $start = 0$, $OPT=\emptyset$, $current$ = $\emptyset$;\\
Compute $E_{T_1}$ and $E_{T_2}$;\\
\While{$i_1< k_1$ or $i_2< k_2$}
{
    \If{$b_1^{i_1}\leq b_2^{i_2}$}
    {
       $\varepsilon_1+=E_{T_1}[i_1++]$;\\
    }
    \Else{
       $\varepsilon_2+=E_{T_2}[i_2++]$ ;\\
    }
    \If{$\varepsilon_1\leq \varepsilon_2$ AND $b_1^{i_1}\geq b_2^{i_2}$}
    {
    		$current = [start, b_1^{i_1}]$;\\
		$OPT \leftarrow OPT \cup \{current\}$;\\        
        $start = b_1^{i_1}+1$;\\
        $\varepsilon_2\leftarrow\varepsilon_1$;
    }
    \If{$\varepsilon_2\leq \varepsilon_1$ AND $b_2^{i_2}\geq b_1^{i_1}$}
    {
    		$current = [start, b_2^{i_2}]$;\\
        $OPT \leftarrow OPT \cup \{current\}$;\\
        $start = b_2^{i_2}+1$;\\
        $\varepsilon_1\leftarrow\varepsilon_2$;
    }
}
Return $OPT$;
\caption{Optimal segment combination (OS)}
\label{alg:optimal_segmentation}
\end{algorithm}

\yannisp{Extremely hard to read. Add small example that extends the example for the GS.}

\yannisp{In algorithm change $i$ and $j$ to $i_1$ and $i_2$ to map cleaner to inputs and other notation. Chunbin:done}

\yannisp{If tight for space the GS should be eliminated both from here and the experimentations. Since OS is provanly optimal there is no reason to promote another technque. At best, we can present in brief (via example) that the interection technique is not tight.}

\subsubsection{Orthogonal projection optimization}
\label{sec:error_guarantee_misalign_LSF}
\rone{D7. Definition of LSF, which is key to the paper is hard to follow. If i understand, i think you mean a family such that for any function f in that family and any translation $a-a'$, there is a function g in that family such that $g(x+a-a') = f(x)$ for all x. If so, definition 2 is a roundabout way. Lack of example, and a mysterious proof in full version make this only more confusing.}

In this part, we present how to apply orthogonal property optimization to improve Formula~4$\textcircled{1}$. Recall that in the aligned case (if the function family is in \textsf{VS}) we can apply the orthogonal property optimization to guarantee $\langle \varepsilon_{T_1^i},\ef{T_2}|_{[a_1^i,b_1^i]}\rangle = 0$. This is because  $\ef{T_2}|_{[a_1^i,b_1^i]} = \ef{T_2^i}$, which is a function in the family. However, in misaligned case $\langle \varepsilon_{T_1^i},\ef{T_2}|_{[a_1^i,b_1^i]}\rangle$ cannot be guaranteed to be $0$ since $\ef{T_2}|_{[a_1^i,b_1^i]}$ may not be a function in the family. For example, in Figure~\ref{fig:segment_combination} $T_2|_{[a_2^1, b_2^1]}$ is not a pre-computed segment in $T_2$, it is just a subsegment. The restriction of the estimation function $\ef{T_2^1}$ to this sub-domain $\ef{T_2}|_{[a_1^1,b_1^1]}$ may not be a function in the family anymore. 

To guarantee the restriction of the function from a bigger domain to a smaller domain is still in the same function family, we identify a function family group called \textsf{linear scalable function family (LSF)}, which is subset of \textsf{VS} but  superset of the polynomial function family.

\smallskip
\noindent\textbf{Linear Scalable Function Family (LSF).}
Informally, a linear scalable family is a function family such that for any function $f$ in that family and any translation $a-a'$, there is a function $f'$ in that family such that $f'(x+a-a') = f(x)$ for all x in the domain.  Definition~\ref{def:linear_scalable_family} gives the formal definition.

\begin{definition} [Linear scalable family (LSF)]
Let $\ff$ be a function family defined in domain $[a,b]$, $\ff$ is a linear scalable family if for any function $f\in\ff$ and any range $[a',b']\subseteq [a,b]$, there exists a function $f'\in\ff$ such that \textsf{Shift}$(f|_{[a',b']},$ $a-a')$ =  $f'|_{[a, a + b' - a']}$. \label{def:linear_scalable_family}
\end{definition}

\begin{lemma}
\label{lemma:polynomial_belongs_LSF}
The polynomial family belongs to the linear scalable family.
\end{lemma}

The proof of Lemma~\ref{lemma:polynomial_belongs_LSF} is shown in Appendix~\ref{appendix:proof_of_polynomial_LSF}.

Recall that, in this paper, we study three different function family groups, i.e.,  ANY, VS, and LSF. Figure~\ref{fig:function_family_xmpls} shows the relation of the three function family groups and also provides example function families for each group.

\eat{
\textcolor{red}{THIS PART GOES TO INTRODUCTION:\\
Figure~\ref{fig:function_family_xmpls}(a) shows the relation of the three function families. Note that  the polynomial family is a subset of the linear scalable family (LSF), which will be formally presented in Lemma~\ref{lemma:polynomial_belongs_LSF}. Figure~\ref{fig:function_family_xmpls}(b) lists some example functions in different families. Those example function families are widely used fitting model functions for time series. For instance,
\cite{pan2017construction} uses polynomial functions to fit time series data, while \cite{tobita2016combined} uses natural logarithmic functions and  natural exponential functions as the fitting model functions for time series data.}}

In the following, we present how to use the orthogonal projection optimization in the misaligned case to improve Formula~4$\textcircled{1}$. 
Let $\bm{f_{T_1}}$ (resp. $\bm{f_{T_2}}$) be the function created from the concatenation of the individual estimation functions on the segments $T_1^i$ $(i\in[1,k_1)$ (resp. $T_2^j$ $(j\in [1,k_2]$)). That is  $\bm{f_{T_1}|_{[a_1^i,b_1^i]}} = \ef{T_1^i}$ for all $i \in [1,k_1]$ and  $\bm{f_{T_2}|_{[a_2^i,b_2^i]}} = \ef{T_2^i}$ for all $i\in[1,k_2]$. \yannisp{Yannis: I think I figured out what was meant to be said. It was not about the bold. It should be bold. Actually there should be more bolds (i.e., vectors). I did a few.} Then the Equation~4$\textcircled{1}$ in the misaligned environment can be reduced as follows. We highlight the parts that would disappear if the segments were aligned.
\begin{align}
&\sum_{i=1}^{k_1}\Big(\fes{T_1^i}\times \overbrace{\|\bm{f_{T_2}|_{[a_1^i,b_1^i]} - f^*_{T_1^i}}\|_2}^{= 0~if~aligned}\Big)\nonumber\\
+&\sum_{i=1}^{k_2}\Big(\fes{T_2^i}\times \overbrace{\|\bm{f_{T_1}|_{[a_2^i,b_2^i]} - f^*_{T_2^i}}\|_2}^{= 0~if~aligned}\Big)~~~~~~~~(5)\nonumber
\end{align}

The proof of the tightness is in Appendix~\ref{appendix:proof_misaligned_times_LSF}. \yannisp{Move to Appendix. We do not have space for the complete proofs. We only have space for intuitions/figures/examples, which we lack here. Also, I do not see the point for separating correctness and tightness proofs. Chunbin: done.}

\noindent\textbf{Efficient Computation of the Error Guarantee}
Notice that both $\|\bm{f_{T_2}|_{[a_1^i,b_1^i]} - f^*_{T_1^i}}\|_2$ and $\|\bm{f_{T_1}|_{[a_2^i,b_2^i]} - f^*_{T_2^i}}\|_2$ can only be computed during query processing time, since only then the pairs of intersecting but misaligned segments become known. A brute force $O(n)$ method, where $n$ is the size of the domain of the segment, would be to literally create the series of $n$ data points predicted by the estimation functions and then perform the straightforward calculation/aggregation described by the formulas. Of course, such brute force approach would require CPU cycles that are proportional to conventional (non-approximate) query processing.
%\footnote{The storage compression advantage of approximate query processing would still be applicable. At any rate, we can also achieve CPU cycle reduction, as we show next.} 
We show that these formulas can be computed in $O(dim(\mathds{F})^3)$ where $dim(\mathds{F})$ is the dimension of the estimation function family. Obviously, the dimension is much smaller than the number of data points in a segment - that is why we employ compression in the first place. For example, for a $1$-degree polynomial function family, $dim(\mathds{F}) = 2$. The key intuition is to store the estimation function's coefficients in an orthonormal basis. The distance between two functions can be efficiently computed using the $dim(\mathds{F})$ coefficients (in the orthonormal basis). Importantly, the orthonormal basis also allows us to compute the coefficients of the restriction of an estimation function in $O(dim(\mathds{F})^3)$. The detailed algorithms and proofs complexity appear in the Appendix~\ref{appendix:computation_formula_5}.

\eat{
We hereby provide an example, while the requisite linear algebra background,

\begin{example}
Consider the polynomial estimation function $5x + 3$ and the orthonormal 2-dimensional basis $(\phi_1, \phi_2)$, where $\bm{\phi_1} = (\frac{1}{\sqrt{2}}, -\frac{1}{\sqrt{2}})$ and $\bm{\phi_2} = (\frac{1}{\sqrt{2}}, \frac{1}{\sqrt{2}})$. Then for the function $f$, Plato stores the coefficients $(\frac{2}{\sqrt{2}}, \frac{8}{\sqrt{2}})$. It does \textit{not} store the coefficients $5$ and $3$, since the vectors $(1,0)$ and $(0,1)$ do not form an orthonormal basis (they are just orthogonal). Nevertheless, it only takes this fast $O(dim(\mathds{F})^2)$ computation to recover the original parameters:
$\frac{2}{\sqrt{2}} \bm{\phi_1} + \frac{8}{\sqrt{2}} \bm{\phi_2} = [5, 3]$

\yannisp{Complete this example. Have two overlapping segments, the one from $T_1$ and the other from $T_2$. The one has the $5x+3$ and the second has its own function. Show the original coordinates. Show the coordinates of the restriction. Show the computation of $\|\bm{f_{T_2}|_{[a_i,b_i]}-\hat{f}^{T_1^i}}\|_2$ and $\|\bm{f_{T_1}|_{[c_i,d_i]}-\hat{f}^{T_2^i}}\|_2$. Chunbin: It is hard to show the restriction, it needs to show the details of the basis transform matrix. I suggest to not provide the example here but in the full version.}
\end{example}
}
\yannisp{Move to the Appendix the rest of the subsection. Chunbin:done.}

%\noindent\textbf{Discussion.}
%If the two time series segmentations are aligned then the first two terms are $0$  and the last term equals to $\sum_{i=1}^k\Big(\|\bm{\varepsilon}_{T_1^i}\|_2\times \|\bm{\varepsilon}_{T_2^i}\|_2\Big)$, which means Equation~\ref{equ:misaligned_LSF} is equal to Equation~\ref{equ:aligned_VS} in the aligned environment.

\noindent\textbf{Elimination of $\ses{T}$.}
If $T$ is compressed by a function in \textsf{LSF}~\footnote{And we know that it many only be combined with other segments compressed by a function in \textsf{LSF}.}, then $\ses{T}$ can be safely eliminated. This is because the error guarantees provided by $\textsf{LSF}$ can get rid of $\ses{T}$ while those given by \textsf{ANY} or \textsf{VS} rely on $\ses{T}$.

\begin{table}[t]
\centering
\begin{tabular}{|c|c|c|c|}
  \hline
  & avg $\#$ of data points & $\#$ of  & \multirow{2}{*}{resolution} \\
                           & in each time series     & time series                       &                             \\ \hline
  HF                       & $126,059,817$           & 15                                & millisecond                 \\ \hline
  HI                       & $2,676,311$             & 14                                & second                      \\ \hline
  HB                       & $1,669,835$             & 16                                & minute                        \\ \hline
  HA                       & $1,587,258$              & 11                               & minute                         \\ \hline
\end{tabular}
\caption{Data Characteristics}
\label{tbl:data_characteristics}
\centering
\begin{tabular}{|c|c|c|}
  \hline
                    & $\#$ of coefficients & $\#$ of error measures \\ \hline
Polynomial & $2$    & $1$\\
Gaussian   & $4$    & $3$\\  \hline
\end{tabular}
\caption{Number of coefficients and error measures}
\label{tbl:number_of_coefficients}
\end{table}

\section{Experiments}
\label{sec:experiments}
\subsection{Environment and Setting}
All experiments were conducted on a computer with a $4^{th}$ Intel $i7$-$4770$ processor (%$4 \times 32$ KB $L1$ data cache, $4\times 256$ KB $L2$ cache, $8$ MB shared $L3$ cache, $4$ physical cores,
$3.6$ GHz), $16$ GB RAM, % and a Seagate $ST2000DM001-1CH1$ hard drive,
running Ubuntu $14.04.1$. 
\eat{
The index building algorithms were implemented in Python using numpy.polynomial. polyfit and scipy.optimize.curve$\_$fit{} packages to find the best fitting linear and non-linear functions. 
}
The algorithms were implemented in $C++$ and were compiled with $g$++ $4.8.4$.

\noindent\textbf{Dataset.}
We evaluated all the error guarantee methods on four real-life datasets:  Historical Forex Data (HF), Historical IoT Data (HI), Historical Bitcoin Exchanges Data (HB), and Historical Air Quality Data (HA). Table~\ref{tbl:data_characteristics} summarizes the data characteristics. The detailed description of each dataset is presented in Appendix~\ref{appendix:experiment}.

\noindent\textbf{Segmentation algorithms.} We adopt the fixed-length segmentation (FL) and the sliding window algorithm (SW). The segments produced by the FL have equal lengths, and will be utilized in our aligned experiments, while the segments created by the SW have variable lengths and are used in our misaligned experiments.

\noindent\textbf{Estimation function families.}
Following the prior work lessons~\cite{keogh1997fast,pan2017construction}, we choose the 1-degree polynomial function family ($\{ax+b|a,b\in R\}$) and the Gaussian function family  ($\{a\exp\Big(\frac{-(x-b)^2}{2c^2}\Big)+d| a, b, c, d\in R\}$) as representatives to compress the time series. Notice that the  Gaussian function family is in ANY, while the polynomial function family is in LSF (also in VS). Table~\ref{tbl:number_of_coefficients} summarizes the number of coefficients and error measures stored for each segment compressed by the corresponding estimation functions.

\smallskip
\noindent\textbf{Queries}
We evaluate the correlation TSA over all the time series pairs in each dataset. The corresponding SQL queries are shown in Appendix~\ref{appendix:experiment}. All the error guarantees and true errors reported in the following are the average values (including the standard variances) across all correlations in a dataset.
%In addition, unless we explain otherwise, we always report the : (i) for queries over aligned time series, if the segments are compressed by functions in LSF (or VS), we use the optimized formula (i.e., Expression~\ref{equ:aligned_VS}) to compute the error guarantees; and (ii) for queries over misaligned time series, we use the optimal segment combination selection strategy (OS) and if the estimation functions are in LSF, then we use the optimized expression (i.e., Expression~\ref{equ:misaligned_LSF}).

\begin{figure}[t]
\center
\includegraphics[width=0.45\textwidth]{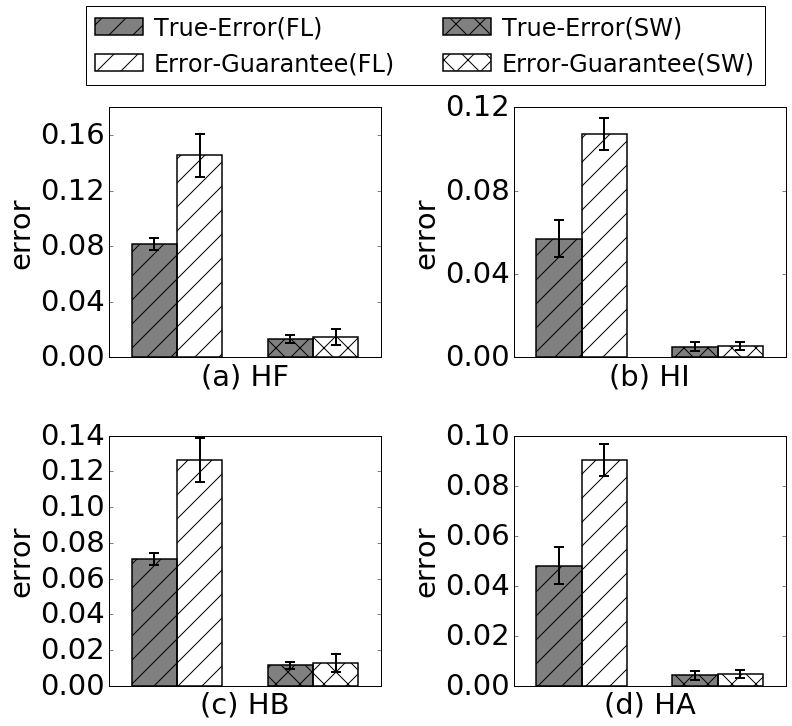}
\vspace{-4mm}
\caption{True errors and error guarantees in aligned (FL)
and misaligned cases (SW). The True-Error(SW) are $0.0132$ and $0.00508$ in (a) and (b).}
\label{fig:align_misalign_HF}
\end{figure}
\subsection{Experimental Results}\label{subsec:experiment}
We evaluate the error guarantees for TSAs over aligned, fixed-length time series segmentations and misaligned, variable-length time series segmentations. In order to provide a fair comparison, we fix the space cost for both cases, i.e., they have the same compression ratios.
%To make sure the segment lists produced by SW and FL have the same compression ratios, we created segment lists as follows: (i) We first run SW for each time series. Let $M$ be the number of segments created by SW; (ii) Then we run FL by setting the segment sizes to be $N/M$ where $N$ is the number of data points in the original time series. In this way, the compression ratios in both cases are $\frac{cM}{N}$ where $c$ is the number of parameters we stored for each segment, i.e., the total number of the coefficients and the error measures.

\noindent \textbf{Error Guarantees Quality}
Figure~\ref{fig:align_misalign_HF} reports the absolute true errors and the error guarantees of the correlation TSAs in the aligned/fixed-length (FL) and misaligned/variable-length (SW) cases using the polynomial function family. Since the TSAs are correlations, the approximate results may range between 1 (perfect correlation) and -1 (perfect reverse correlation), with 0 meaning no correlation at all.

Under the same compression ratio~\footnote{Compression ratio is the size of the original data over the size of the compressed data.} 
the variable-length error guarantees are much smaller than the fixed-length error guarantees. In Figure~\ref{fig:align_misalign_HF}, the misaligned Error-Guarantee (SW)  is $10\times\sim20\times$ smaller than the aligned Error-Guarantee (FL) on the average (ranging the compression ratio from $10,000$ to $100$). This is mainly because, as it has already been known, variable-length allows for much better estimation. Indeed, notice the misaligned true errors are also much smaller than the aligned true errors. For example, In Figure~\ref{fig:align_misalign_HF}, True-Error(SW) is $6\times\sim11\times$ smaller than True-Error(FL)  on the average.

Importantly, the error guarantees are close to the true errors, especially for the misaligned error guarantees, which matter most practically. In particular, Error-Guarantee(SW) is only $1.08\times\sim 1.11\times$ larger than the True-Error(SW) in HF and HI respectively (on the average). Furthermore, they are very small in absolute terms. This indicates the high quality and practicality of AI (Amplitude-independent) error guarantees.

\begin{figure}[t]
\center
\includegraphics[width=0.45\textwidth]{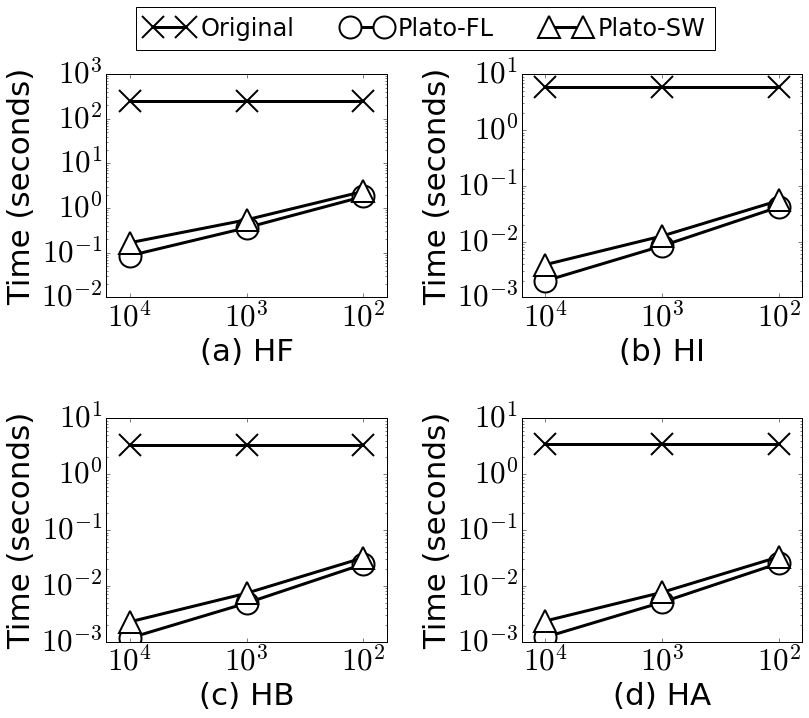}
\vspace{-4mm}
\caption{Running time of TSAs in aligned and misaligned cases.}
\label{fig:align_misalign_time}
\end{figure}

\noindent \textbf{Run time performance}
Figure~\ref{fig:align_misalign_time} reports the total running time of the correlation TSAs over (i) the original time series (Original), (ii) the time series segmented into a fixed length, aligned segments  (Plato-FL) and (iii) time series segmented into misaligned, variable-length segments by SW (Plato-SW). The estimation function family is the polynomial family. The x-axis is the compression ratio (from $10000$ to $100$).

Both Plato-FL and Plato-SW outperform vastly the Original in all the datasets. For example, when the compression ratio is $1000$, Plato-FL and Plato-SW are about three orders of magnitude faster than Original.

Plato-SW is about $1.8\times$ slower than Plato-FL due to the intricacy of the segment combination selection algorithm. However, a mere 80\% penalty is a minor price to pay for the orders-of-magnitude superior error guarantees delivered by misaligned/variable-length segmentations.

\begin{figure}[t]
	\begin{center}
		\includegraphics[width=0.4\textwidth]{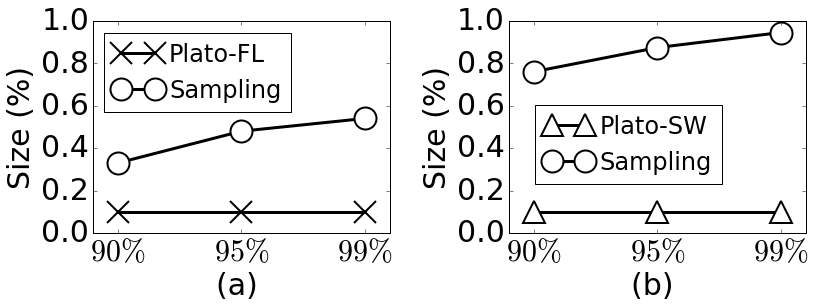}
	\end{center}
	\vspace{-4mm}
	\caption{Space cost of sampling and \db{} when providing the same error guarantees.}
	\label{fig:sampling}
	\begin{center}
		\includegraphics[width=0.4\textwidth]{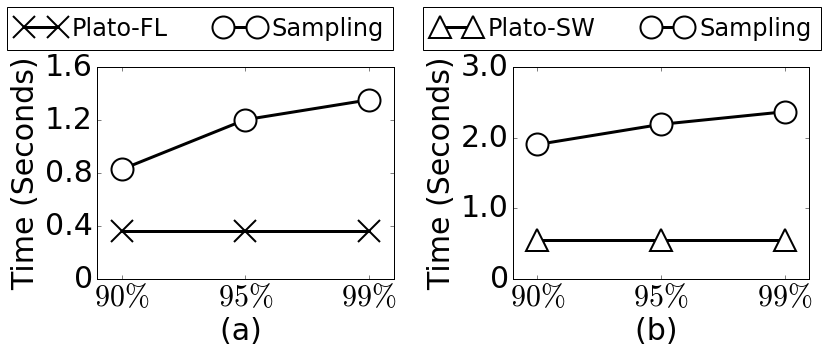}	
	\end{center}
	\vspace{-4mm}
	\caption{Running time of sampling and \db{} when providing the same error guarantees.}
	\label{fig:sampling_time}
\end{figure}

\noindent \textbf{Comparison with sampling}
In this part, we compare (i) the space cost and (ii) the runtime performance of \db{} with the sampling methods when providing similar error guarantees. We use a uniform random sampling scheme with a global seed in order to create a samples database. We also assume knowledge of minimums and maximums. That is, let $X_1,...,X_n$ be the random variables such that $d_{min}\leq X_i \leq d_{max}$ for all $i$ where $X_i = d^{T_1}_i\times d^{T_2}_i$, $d_{min} = \min\{d^{T_1}_i\}\times \min\{d^{T_2}_i\}$, and  $d_{max} = \max\{d^{T_1}_i\}\times \max\{d^{T_2}_i\}$. Let $R = \sum_{i=1}^nX_i$ and $\varepsilon$ be the error guarantee. Using the
%Hoeffding bounds~\cite{hoeffding1963probability} and
Chernoff bounds~\cite{hagerup1990guided},
%we have:
%\begin{align*}
%\Pr[|R-E[R]|\leq \varepsilon] \geq 1- 2\exp\Big(-\frac{2n\varepsilon^2}{(d_{max} - d_{min})^2}\Big)
%\end{align*}
%Based on this formula,
we can obtain the minimal sample size needed in order to achieve the desired error guarantee with certain confidence.

\begin{figure}[t]
\centering
\includegraphics[width=0.4\textwidth]{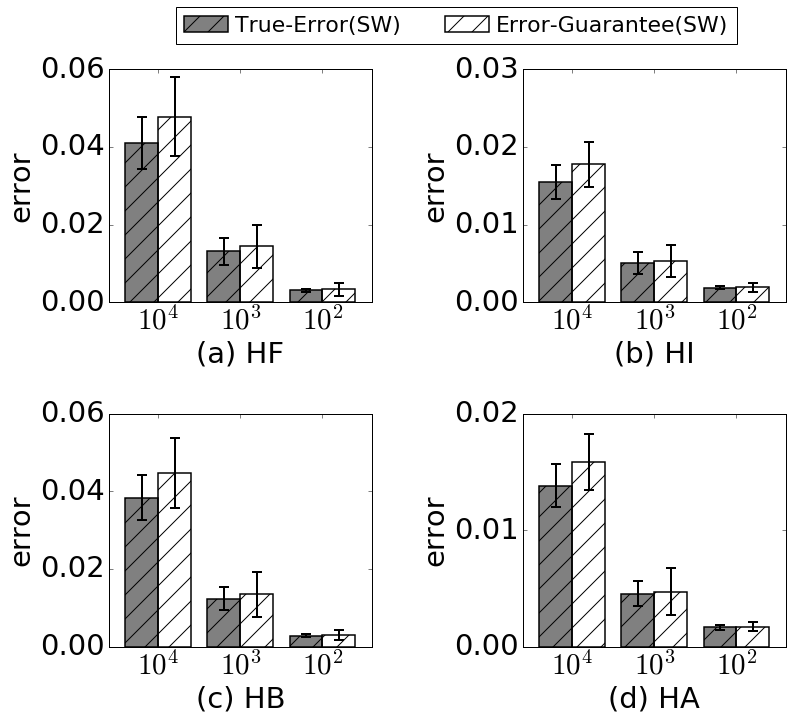}
\vspace{-4mm}
\caption{Effect of compression ratios.}
\label{fig:compression_ratio}
\centering
\includegraphics[width=0.4\textwidth]{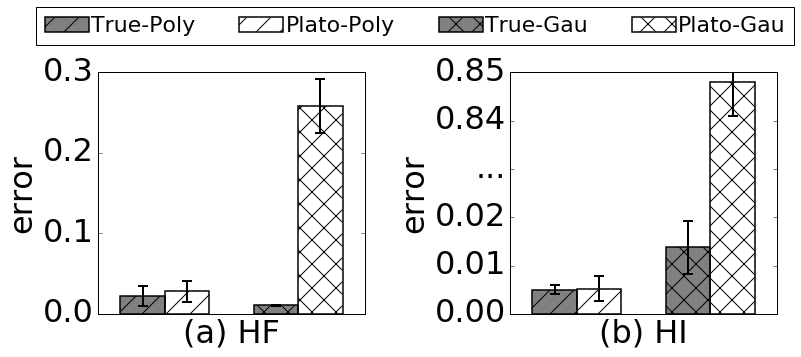}
\vspace{-4mm}
\caption{Effect of estimation function families.}
\label{fig:function_family}
\end{figure}

Figure~\ref{fig:sampling} reports the sizes (as percentage to the original data size) of sampled data points in order to provide similar error guarantees with the Plato-FL (the error guarantee of TSAs over aligned, fixed-length time series produced by FL) and Plato-SW  (the error guarantee of TSAs over misaligned time series produced by SW) with $1000$ compression ratio in HF respectively.  Figure~\ref{fig:sampling_time} shows the corresponding runtime cost. To achieve similar error guarantees, sampling needs more space and more time than \db.  We define ``similar" to mean 90\%, or 95\% or 99\% confidence - in contrast to Plato's deterministic, 100\% confidence guarantees.
\eat{
\noindent\textbf{Additional experiments.} We evaluate the effects of individual factors in Appendix~\ref{exp:effect}. More precisely, we study the effects of (i) compression ratios, (ii) estimation function families, (iii) orthogonal optimizations, and (iv) segment combination selection strategies.
}

%\begin{figure}[t]
%\centering
%\includegraphics[width=0.48\textwidth]{Experiment/index-building-time.png}
%\vspace{-7mm}
%\caption{Segment list index building time.}
%\label{fig:index_building_time}
%\end{figure}

\subsubsection{Effects of Individual Factors}
\label{exp:effect}

In this part, we study the effects of (i) compression ratios, (ii) estimation function families, (iii) orthogonal optimizations, and (iv) segment combination selection strategies.

\noindent \textbf{Compression ratios}
%\label{exp:compression_ratio}
In order to isolate the effect of the compression ratios,~\footnote{Compression ratio is the size of the original data over the size of the compressed data.} we fix the estimation function family to be polynomials and fix the segment list building algorithm to be SW. In Figure~\ref{fig:compression_ratio}, we change the compression ratios from $10,000$ to $100$ by controlling the error threshold values and report the corresponding true errors (True-Error(SW)) and the error guarantees (Error-Guarantee(SW)).

Naturally, higher compression ratios lead to smaller true errors and error guarantees. For example, in Figure~\ref{fig:compression_ratio}(a), the true error and error guarantee with $100$ compression ratio are $13.32\times$ and $15.58\times$ smaller than those with $10,000$ compression ratio on the average.
Importantly, the error guarantees provided by \db{} are close to the true error in all the datasets and are generally small in absolute terms (with the relative exception of $10,000$ compression on HF). Again, this indicates the high quality of the error guarantees provided by \db.

\noindent \textbf{Estimation function families}
In order to isolate the effect of the  estimation function families, we fix the segment list building algorithm to be SW and fix the compression ratio to $1000$. Figure~\ref{fig:function_family} presents the true errors and the error guarantees for TSAs over time series compressed by polynomial functions (True-Error(Poly), Error-Guarantee(Poly)) and Gaussian functions (True-Error(Gau), Error-Guarantee(Gau)) respectively.

The error guarantees with estimation functions from LSF (polynomials) are significantly smaller than those with estimation functions in ANY (Gaussians). In Figure~\ref{fig:function_family}(a),   Error-Guarantee(Poly) (in LSF and VS) is about $10\times$ smaller than  Error-Guarantee(Gau) (in ANY)  on the average and in Figure~\ref{fig:function_family}(b),  Error-Guarantee(Poly) (in LSF and VS) is about $160\times$ smaller than  Error-Guarantee(Gau) (in ANY)  on the average. Notice that the error guarantees provided by Plato-Poly is AI, while those of Plato-Gau are not. So the results show that AI error guarantees are practical while non-AI error guarantees are not. Interestingly, True-Error(Gau) is smaller than True-Error(Poly) in the HF dataset, which indicates that Gaussian functions model HF data better than the polynomial functions - not surprising given the more random movements of financial data. The guarantees produced by the polynomials are far better thanks to AI.

\begin{figure}[t]
\centering
\includegraphics[width=0.42\textwidth]{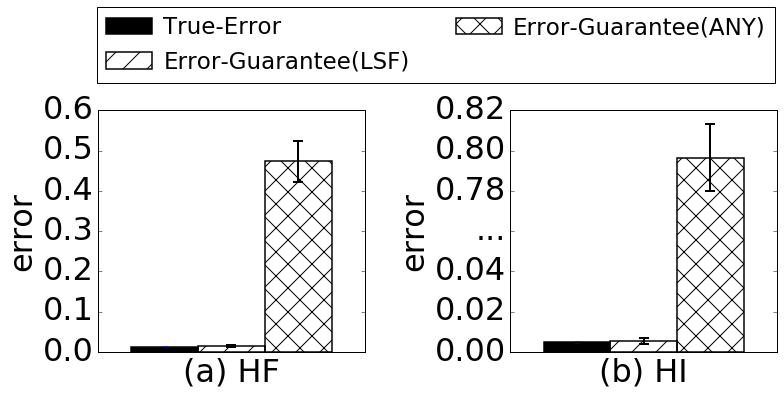}
\vspace{-5mm}
\caption{Effect of orthogonal optimization.}
\label{fig:orthogonal_optimization}
\centering
\includegraphics[width=0.42\textwidth]{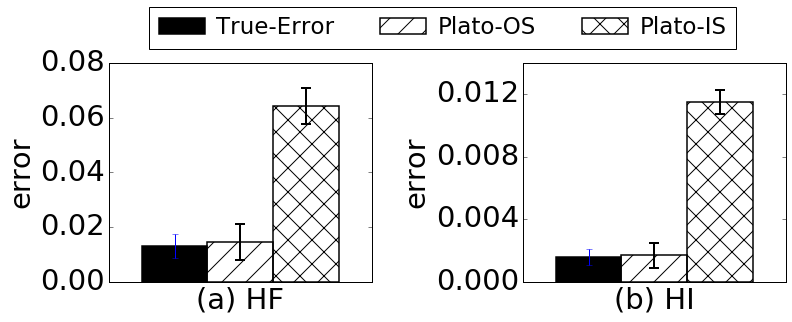}
\vspace{-5mm}
\caption{Effect of segment combination selection strategies.}
\label{fig:combination_selection}
\end{figure}

\noindent \textbf{Effect of Orthogonal Optimization and LSF}
To measure the effect on error guarantees of the orthogonal optimization (and its extension to misaligned segmentations, enabled by LSF) we fix the estimation function family to the polynomials, which are LSF and, trivially, are also in ANY. We use both the general error guarantees of ANY (Error-Guarantee(ANY)) and the specialized error guarantees of LSF (Error-Guarantee(LSF)) for TSAs over misaligned segments compressed by polynomial functions (using variable-length segmentations with the SW algorithm). We fix the compression ratio to $1000$. As shown in Figure~\ref{fig:orthogonal_optimization}, the error guarantee for LSF certifies that the true result is just within $\pm0.0137$ \yannisp{plus/minus X - mention the HF absolute value X, it is so small that cannot be understood from the figure. Chunbin:done.} in HF and within $\pm0.0052$\yannisp{plus/minus Y. Chunbin:done.} in HI.

\noindent \textbf{Segment combination selection strategies}
To isolate the quality effect of employing the optimal segment combination selection strategy (OS) we compare it with IS strategy (the straightforwad method mentioned in Section~\ref{sec:segment_combination_selection}) on a case of variable-length compression with an LSF function family (polynomials).
Figure~\ref{fig:combination_selection} shows that Plato-OS is about $5\times$ smaller than Plato-MS on the average. In addition, the running time of Plato-IS and Plato-OS are close. For example, the running time of Plato-IS and Plato-OS are $0.536$ and $0.548$ seconds in HF respectively.

\vspace{-2mm}
\section{Related Work}
\label{sec:related_work}
%\yannisk{Chunbin, please read section below and address the comments. While reading, please make sure that everything we state is true (especially the absolute comments of this being the first work that...). Otherwise, please leave a comment.}
Approximate query processing (AQP) and data compression have been widely studied.  %the focus on an extensive body of work
whose most relevant aspects are summarized next. 
%To the best of our knowledge, this is the first work that provides deterministic guarantees for analytics over multiple compressed time series.

\noindent\textbf{AQP with probabilistic error guarantees.} Approximate query processing using \emph{sampling}~\cite{chaudhuri2007optimized,SidirourgosKB11,PansareBJC11, AgarwalMPMMS13} computes approximate answers by appropriately evaluating the queries on small samples of the data, e.g., STRAT~\cite{chaudhuri2007optimized}, SciBORQ~\cite{SidirourgosKB11}, and BlinkDB~\cite{AgarwalMPMMS13}. Such approaches typically leverage statistical inequalities and the central limit theorem to compute the confidence interval (or variance) of the computed approximate answer. As a result, their error guarantees are probabilistic - as opposed to this work's deterministic (100\% confidence) ones. Note however that, unlike sampling, our compression-based techniques are tuned for time series and continuous data.

\noindent\textbf{AQP with deterministic error guarantees.} Approximately answering queries while providing deterministic error guarantees has been successfully applied in many applications~\cite{DBLP:conf/icde/CormodeKMS05,DBLP:conf/sigmod/GreenwaldK01,DBLP:conf/sigmod/RajagopalanML98,PottiP15,LazaridisM01,PoosalaIHS96}. However, existing work in the area has focused on simple aggregation queries that involve only a single time series (or table) and aggregates such as SUM, COUNT, MIN, MAX and AVG. Our work extends the prior work, as it addresses analytics over multiple compressed time series such as correlation, cross-correlation. In addition, this work is the first one to categorize compression function families based on their suitability for error guarantees.

%\noindent\textbf{Approximate query answering over sensor data.} Moreover, \db\ is one of the first approximate query answering systems that leverage the fact that sensor data are not random but follow a usually smooth underlying phenomenon. The majority of existing works on approximate query answering looked at general relational data. Moreover, the ones that studied approximate query processing for sensor data, focused on the networking aspect of the problem, studying how aggregate queries can be efficiently evaluated in a distributed sensor network~\cite{madden2002tag,considine2004approximate,ConsidineLKB04}.
%While these works focused on the networking aspect of sensor data, our work focuses on the continuous nature of the sensor data, which it leverages to accelerate query processing even in a single machine scenario, where historical sensor data already accumulated on the machine have to be analyzed.\\

\noindent \textbf{Data summarizations and compressions}
%Relevant work in this area has come mostly from two different communities: the database community~\cite{IoannidisP95,PoosalaIHS96,PapapetrouGD12,Ting16}
%and the signal processing community~\cite{KeoghCPM01,keogh2001locally,keogh1997fast,ChanF99,FaloutsosRM94}.
The database community has mostly focused on creating summarizations (also referred to as synopses or sketches) that can be used to answer specific queries. These include among others histograms 
%\cite{IoannidisP95,PoosalaIHS96,GibbonsMP97,PoosalaGI99} 
\cite{ShapiroC84,IoannidisP95,WangS08,ReissGH06}
(e.g., EquiWidth and EquiDepth histograms \cite{ShapiroC84}, V-Optimal histograms~\cite{IoannidisP95}, and Hierarchical Model Fitting (HMF) histograms~\cite{WangS08}), 
\yannisp{THIS SHOULD MOVE TO SAMPLING: as well as sampling methods~\cite{HaasS92,ChenY17},} used among other for cardinality estimation~\cite{IoannidisP95} and selectivity estimation~\cite{PoosalaIHS96}.
%Several types of histograms have been proposed and evaluated experimentally in terms of their accuracy, including EquiWidth and EquiHeight [Koo80,SC84], MaxDiff, V-Optimal histograms [IP95, PIHS96]. A formal taxonomy of histograms was proposed in [PIHS96]. The V-Optimal histograms have been shown to minimize the average error for several selectivity estimation problems
%
%In contrast to such special-purpose approaches, we support a large class of queries over arbitrary time series.
The signal processing community  produced a variety of methods that can be used to compress time series data and thus are more relevant to the present work, as they provide the underlying compressions. These include among others the Piecewise Aggregate Approximation (PAA)~\cite{KeoghCPM01}, and %the  Adaptive Piecewise Constant Approximation (APCA)~\cite{keogh2001locally},
the Piecewise Linear Representation (PLR)~\cite{keogh1997fast}.
%, the Discrete Wavelet Transform (DWT)~\cite{ChanF99}, and the Discrete Fourier Transform (DFT)~\cite{FaloutsosRM94}.
\db{} is orthogonal to those data summarization and compression techniques.

%\section{Closing Remarks and Future Work}

\section{Summary and Future Direction}
\label{sec:summary_future}
This work indicates that deterministic error guarantees are feasible and practical, given the appropriate combination of error measures and estimation function family. Future work may develop such combinations for other important families also. Note that the tightness results of this paper do not preclude the future development of practical and theoretically-sound deterministic error guarantees for families are currently outside the LSF (or outside the VS in the case of aligned series). 
%Rather, the tightness results merely state that (i) without introducing additional assumptions about a function family and/or (ii) without introducing different/additional error measures, no better error guarantees can be achieved. 
Researchers may come up with other interesting properties of function families outside LSF (or VS) and deliver good error guarantees, based on such properties.  
\eat{
Finally, if some cases turn out to be unaddressable by deterministic error guarantees, another interesting area of research would be the delivery of probabilistic guarantees over compressed data.

While this work focused on basic vector operations, a number of applications-oriented adjustments and extensions appear possible. In such cases the results of this work will become building blocks in more complex analysis settings. One such case is analyses that correspond to multiple queries (in contrast to this fundamentals-oriented work that considered queries that return a single number). For example, in an anomaly detection setting we would not simply be asking how much cross-correlated is the air conditioning air flow with the air temperature. 
%We would rather ask for the time periods when the temperature decorrelated significantly from the air flow. 
It is easy to see how the latter can be framed as multiple queries. Of course, additional optimizations will be possible in the multiple query processing case.
}

\balance{}

\newpage

\bibliographystyle{ACM-Reference-Format}
%\bibliography{sigmod.bib}
\bibliography{ref/References.bib,ref/deterministic_error_guarantee.bib,ref/time_series_normalization.bib,ref/time_series_compression.bib}


\begin{thebibliography}{47}

%%% the Web address.

\ifx \showCODEN    \undefined \def \showCODEN     #1{\unskip}     \fi
\ifx \showDOI      \undefined \def \showDOI       #1{#1}\fi
\ifx \showISBNx    \undefined \def \showISBNx     #1{\unskip}     \fi
\ifx \showISBNxiii \undefined \def \showISBNxiii  #1{\unskip}     \fi
\ifx \showISSN     \undefined \def \showISSN      #1{\unskip}     \fi
\ifx \showLCCN     \undefined \def \showLCCN      #1{\unskip}     \fi
\ifx \shownote     \undefined \def \shownote      #1{#1}          \fi
\ifx \showarticletitle \undefined \def \showarticletitle #1{#1}   \fi
\ifx \showURL      \undefined \def \showURL       {\relax}        \fi
% invisible to TeX
\providecommand\bibfield[2]{#2}
\providecommand\bibinfo[2]{#2}
\providecommand\natexlab[1]{#1}
\providecommand\showeprint[2][]{arXiv:#2}

\bibitem[\protect\citeauthoryear{Agarwal, Mozafari, Panda, Milner, Madden, and
  Stoica}{Agarwal et~al\mbox{.}}{2013}]%
        {AgarwalMPMMS13}
\bibfield{author}{\bibinfo{person}{Sameer Agarwal}, \bibinfo{person}{Barzan
  Mozafari}, \bibinfo{person}{Aurojit Panda}, \bibinfo{person}{Henry Milner},
  \bibinfo{person}{Samuel Madden}, {and} \bibinfo{person}{Ion Stoica}.}
  \bibinfo{year}{2013}\natexlab{}.
\newblock \showarticletitle{BlinkDB: queries with bounded errors and bounded
  response times on very large data}. In \bibinfo{booktitle}{\emph{EuroSys}}.
  \bibinfo{pages}{29--42}.
\newblock


\bibitem[\protect\citeauthoryear{Aghabozorgi, Shirkhorshidi, and
  Teh}{Aghabozorgi et~al\mbox{.}}{2015}]%
        {journals/is/AghabozorgiST15}
\bibfield{author}{\bibinfo{person}{Saeed~Reza Aghabozorgi},
  \bibinfo{person}{Ali~Seyed Shirkhorshidi}, {and} \bibinfo{person}{Ying~Wah
  Teh}.} \bibinfo{year}{2015}\natexlab{}.
\newblock \showarticletitle{Time-series clustering - {A} decade review}.
\newblock \bibinfo{journal}{\emph{Inf. Syst.}}  \bibinfo{volume}{53}
  (\bibinfo{year}{2015}), \bibinfo{pages}{16--38}.
\newblock


\bibitem[\protect\citeauthoryear{Chan and Fu}{Chan and Fu}{1999}]%
        {ChanF99}
\bibfield{author}{\bibinfo{person}{Kin{-}pong Chan} {and}
  \bibinfo{person}{Ada~Wai{-}Chee Fu}.} \bibinfo{year}{1999}\natexlab{}.
\newblock \showarticletitle{Efficient Time Series Matching by Wavelets}. In
  \bibinfo{booktitle}{\emph{ICDE}}. \bibinfo{pages}{126--133}.
\newblock


\bibitem[\protect\citeauthoryear{Chaudhuri, Das, and Narasayya}{Chaudhuri
  et~al\mbox{.}}{2007}]%
        {chaudhuri2007optimized}
\bibfield{author}{\bibinfo{person}{Surajit Chaudhuri}, \bibinfo{person}{Gautam
  Das}, {and} \bibinfo{person}{Vivek Narasayya}.}
  \bibinfo{year}{2007}\natexlab{}.
\newblock \showarticletitle{Optimized stratified sampling for approximate query
  processing}.
\newblock \bibinfo{journal}{\emph{TODS}} \bibinfo{volume}{32},
  \bibinfo{number}{2} (\bibinfo{year}{2007}), \bibinfo{pages}{9}.
\newblock


\bibitem[\protect\citeauthoryear{Chaudhuri, Ding, and Kandula}{Chaudhuri
  et~al\mbox{.}}{2017}]%
        {chaudhuri2017approximate}
\bibfield{author}{\bibinfo{person}{Surajit Chaudhuri}, \bibinfo{person}{Bolin
  Ding}, {and} \bibinfo{person}{Srikanth Kandula}.}
  \bibinfo{year}{2017}\natexlab{}.
\newblock \showarticletitle{Approximate query processing: no silver bullet}. In
  \bibinfo{booktitle}{\emph{Sigmod}}. ACM, \bibinfo{pages}{511--519}.
\newblock


\bibitem[\protect\citeauthoryear{Chen and Ng}{Chen and Ng}{2004}]%
        {DBLP:conf/vldb/ChenN04}
\bibfield{author}{\bibinfo{person}{Lei Chen} {and} \bibinfo{person}{Raymond~T.
  Ng}.} \bibinfo{year}{2004}\natexlab{}.
\newblock \showarticletitle{On The Marriage of Lp-norms and Edit Distance}. In
  \bibinfo{booktitle}{\emph{{VLDB}}}. \bibinfo{pages}{792--803}.
\newblock


\bibitem[\protect\citeauthoryear{Cheney and Kincaid}{Cheney and
  Kincaid}{2009}]%
        {cheney2009linear}
\bibfield{author}{\bibinfo{person}{Ward Cheney} {and} \bibinfo{person}{David
  Kincaid}.} \bibinfo{year}{2009}\natexlab{}.
\newblock \showarticletitle{Linear algebra: Theory and applications}.
\newblock \bibinfo{journal}{\emph{The Australian Mathematical Society}}
  \bibinfo{volume}{110} (\bibinfo{year}{2009}).
\newblock


\bibitem[\protect\citeauthoryear{Choi}{Choi}{2012}]%
        {choi2012arma}
\bibfield{author}{\bibinfo{person}{ByoungSeon Choi}.}
  \bibinfo{year}{2012}\natexlab{}.
\newblock \bibinfo{booktitle}{\emph{ARMA model identification}}.
\newblock \bibinfo{publisher}{Springer Science \& Business Media}.
\newblock


\bibitem[\protect\citeauthoryear{Cormode, Korn, Muthukrishnan, and
  Srivastava}{Cormode et~al\mbox{.}}{2005}]%
        {DBLP:conf/icde/CormodeKMS05}
\bibfield{author}{\bibinfo{person}{Graham Cormode}, \bibinfo{person}{Flip
  Korn}, \bibinfo{person}{S. Muthukrishnan}, {and} \bibinfo{person}{Divesh
  Srivastava}.} \bibinfo{year}{2005}\natexlab{}.
\newblock \showarticletitle{Effective Computation of Biased Quantiles over Data
  Streams}. In \bibinfo{booktitle}{\emph{{ICDE}}}. \bibinfo{pages}{20--31}.
\newblock


\bibitem[\protect\citeauthoryear{Denison, Mallick, and Smith}{Denison
  et~al\mbox{.}}{1998}]%
        {denison1998automatic}
\bibfield{author}{\bibinfo{person}{DGT Denison}, \bibinfo{person}{BK Mallick},
  {and} \bibinfo{person}{AFM Smith}.} \bibinfo{year}{1998}\natexlab{}.
\newblock \showarticletitle{Automatic Bayesian curve fitting}.
\newblock \bibinfo{journal}{\emph{Journal of the Royal Statistical Society:
  Series B (Statistical Methodology)}} \bibinfo{volume}{60},
  \bibinfo{number}{2} (\bibinfo{year}{1998}), \bibinfo{pages}{333--350}.
\newblock


\bibitem[\protect\citeauthoryear{Eisenberg and Melton}{Eisenberg and
  Melton}{2002}]%
        {eisenberg2002sql}
\bibfield{author}{\bibinfo{person}{Andrew Eisenberg} {and} \bibinfo{person}{Jim
  Melton}.} \bibinfo{year}{2002}\natexlab{}.
\newblock \showarticletitle{SQL/XML is making good progress}.
\newblock \bibinfo{journal}{\emph{ACM Sigmod Record}} \bibinfo{volume}{31},
  \bibinfo{number}{2} (\bibinfo{year}{2002}), \bibinfo{pages}{101--108}.
\newblock


\bibitem[\protect\citeauthoryear{Faloutsos, Ranganathan, and
  Manolopoulos}{Faloutsos et~al\mbox{.}}{1994}]%
        {FaloutsosRM94}
\bibfield{author}{\bibinfo{person}{Christos Faloutsos}, \bibinfo{person}{M.
  Ranganathan}, {and} \bibinfo{person}{Yannis Manolopoulos}.}
  \bibinfo{year}{1994}\natexlab{}.
\newblock \showarticletitle{Fast Subsequence Matching in Time-Series
  Databases}. In \bibinfo{booktitle}{\emph{SIGMOD}}. \bibinfo{pages}{419--429}.
\newblock


\bibitem[\protect\citeauthoryear{Galakatos, Crotty, Zgraggen, Binnig, and
  Kraska}{Galakatos et~al\mbox{.}}{2017}]%
        {galakatos2017revisiting}
\bibfield{author}{\bibinfo{person}{Alex Galakatos}, \bibinfo{person}{Andrew
  Crotty}, \bibinfo{person}{Emanuel Zgraggen}, \bibinfo{person}{Carsten
  Binnig}, {and} \bibinfo{person}{Tim Kraska}.}
  \bibinfo{year}{2017}\natexlab{}.
\newblock \showarticletitle{Revisiting reuse for approximate query processing}.
\newblock \bibinfo{journal}{\emph{Proceedings of the VLDB Endowment}}
  \bibinfo{volume}{10}, \bibinfo{number}{10} (\bibinfo{year}{2017}),
  \bibinfo{pages}{1142--1153}.
\newblock


\bibitem[\protect\citeauthoryear{Greenwald and Khanna}{Greenwald and
  Khanna}{2001}]%
        {DBLP:conf/sigmod/GreenwaldK01}
\bibfield{author}{\bibinfo{person}{Michael Greenwald} {and}
  \bibinfo{person}{Sanjeev Khanna}.} \bibinfo{year}{2001}\natexlab{}.
\newblock \showarticletitle{Space-Efficient Online Computation of Quantile
  Summaries}. In \bibinfo{booktitle}{\emph{{SIGMOD}}}. \bibinfo{pages}{58--66}.
\newblock


\bibitem[\protect\citeauthoryear{Hagerup and R{\"u}b}{Hagerup and
  R{\"u}b}{1990}]%
        {hagerup1990guided}
\bibfield{author}{\bibinfo{person}{Torben Hagerup} {and}
  \bibinfo{person}{Christine R{\"u}b}.} \bibinfo{year}{1990}\natexlab{}.
\newblock \showarticletitle{A guided tour of Chernoff bounds}.
\newblock \bibinfo{journal}{\emph{Information processing letters}}
  \bibinfo{volume}{33}, \bibinfo{number}{6} (\bibinfo{year}{1990}),
  \bibinfo{pages}{305--308}.
\newblock


\bibitem[\protect\citeauthoryear{Hoffmann}{Hoffmann}{1989}]%
        {hoffmann1989iterative}
\bibfield{author}{\bibinfo{person}{Walter Hoffmann}.}
  \bibinfo{year}{1989}\natexlab{}.
\newblock \showarticletitle{Iterative algorithms for Gram-Schmidt
  orthogonalization}.
\newblock \bibinfo{journal}{\emph{Computing}} \bibinfo{volume}{41},
  \bibinfo{number}{4} (\bibinfo{year}{1989}), \bibinfo{pages}{335--348}.
\newblock


\bibitem[\protect\citeauthoryear{Ioannidis and Poosala}{Ioannidis and
  Poosala}{1995}]%
        {IoannidisP95}
\bibfield{author}{\bibinfo{person}{Yannis~E. Ioannidis} {and}
  \bibinfo{person}{Viswanath Poosala}.} \bibinfo{year}{1995}\natexlab{}.
\newblock \showarticletitle{Balancing Histogram Optimality and Practicality for
  Query Result Size Estimation}. In \bibinfo{booktitle}{\emph{SIGMOD}}.
  \bibinfo{pages}{233--244}.
\newblock


\bibitem[\protect\citeauthoryear{Keogh}{Keogh}{1997}]%
        {keogh1997fast}
\bibfield{author}{\bibinfo{person}{Eamonn Keogh}.}
  \bibinfo{year}{1997}\natexlab{}.
\newblock \showarticletitle{Fast similarity search in the presence of
  longitudinal scaling in time series databases}. In
  \bibinfo{booktitle}{\emph{ICTAI}}. \bibinfo{pages}{578--584}.
\newblock


\bibitem[\protect\citeauthoryear{Keogh, Chu, Hart, and Pazzani}{Keogh
  et~al\mbox{.}}{2001b}]%
        {keogh2001online}
\bibfield{author}{\bibinfo{person}{Eamonn Keogh}, \bibinfo{person}{Selina Chu},
  \bibinfo{person}{David Hart}, {and} \bibinfo{person}{Michael Pazzani}.}
  \bibinfo{year}{2001}\natexlab{b}.
\newblock \showarticletitle{An online algorithm for segmenting time series}. In
  \bibinfo{booktitle}{\emph{ICDM}}. \bibinfo{pages}{289--296}.
\newblock


\bibitem[\protect\citeauthoryear{Keogh, Chu, Hart, and Pazzani}{Keogh
  et~al\mbox{.}}{2004}]%
        {keogh2004segmenting}
\bibfield{author}{\bibinfo{person}{Eamonn Keogh}, \bibinfo{person}{Selina Chu},
  \bibinfo{person}{David Hart}, {and} \bibinfo{person}{Michael Pazzani}.}
  \bibinfo{year}{2004}\natexlab{}.
\newblock \showarticletitle{Segmenting time series: A survey and novel
  approach}.
\newblock In \bibinfo{booktitle}{\emph{Data mining in time series databases}}.
  \bibinfo{publisher}{World Scientific}, \bibinfo{pages}{1--21}.
\newblock


\bibitem[\protect\citeauthoryear{Keogh, Chakrabarti, Pazzani, and
  Mehrotra}{Keogh et~al\mbox{.}}{2001a}]%
        {KeoghCPM01}
\bibfield{author}{\bibinfo{person}{Eamonn~J. Keogh}, \bibinfo{person}{Kaushik
  Chakrabarti}, \bibinfo{person}{Michael~J. Pazzani}, {and}
  \bibinfo{person}{Sharad Mehrotra}.} \bibinfo{year}{2001}\natexlab{a}.
\newblock \showarticletitle{Dimensionality Reduction for Fast Similarity Search
  in Large Time Series Databases}.
\newblock \bibinfo{journal}{\emph{KAIS}} \bibinfo{volume}{3},
  \bibinfo{number}{3} (\bibinfo{year}{2001}), \bibinfo{pages}{263--286}.
\newblock


\bibitem[\protect\citeauthoryear{Keogh and Pazzani}{Keogh and Pazzani}{1998}]%
        {conf/kdd/KeoghP98}
\bibfield{author}{\bibinfo{person}{Eamonn~J. Keogh} {and}
  \bibinfo{person}{Michael~J. Pazzani}.} \bibinfo{year}{1998}\natexlab{}.
\newblock \showarticletitle{An Enhanced Representation of Time Series Which
  Allows Fast and Accurate Classification, Clustering and Relevance Feedback}.
  In \bibinfo{booktitle}{\emph{KDD}}. \bibinfo{pages}{239--243}.
\newblock


\bibitem[\protect\citeauthoryear{Keogh and Pazzani}{Keogh and Pazzani}{1999}]%
        {keogh1999relevance}
\bibfield{author}{\bibinfo{person}{Eamonn~J Keogh} {and}
  \bibinfo{person}{Michael~J Pazzani}.} \bibinfo{year}{1999}\natexlab{}.
\newblock \showarticletitle{Relevance feedback retrieval of time series data}.
  In \bibinfo{booktitle}{\emph{SIGIR}}. \bibinfo{pages}{183--190}.
\newblock


\bibitem[\protect\citeauthoryear{Kim}{Kim}{2003}]%
        {Kim03}
\bibfield{author}{\bibinfo{person}{Kyoung{-}jae Kim}.}
  \bibinfo{year}{2003}\natexlab{}.
\newblock \showarticletitle{Financial time series forecasting using support
  vector machines}.
\newblock \bibinfo{journal}{\emph{Neurocomputing}} \bibinfo{volume}{55},
  \bibinfo{number}{1-2} (\bibinfo{year}{2003}), \bibinfo{pages}{307--319}.
\newblock


\bibitem[\protect\citeauthoryear{Koski, Juhola, and Meriste}{Koski
  et~al\mbox{.}}{1995}]%
        {koski1995syntactic}
\bibfield{author}{\bibinfo{person}{Antti Koski}, \bibinfo{person}{Martti
  Juhola}, {and} \bibinfo{person}{Merik Meriste}.}
  \bibinfo{year}{1995}\natexlab{}.
\newblock \showarticletitle{Syntactic recognition of ECG signals by attributed
  finite automata}.
\newblock \bibinfo{journal}{\emph{Pattern Recognition}} \bibinfo{volume}{28},
  \bibinfo{number}{12} (\bibinfo{year}{1995}), \bibinfo{pages}{1927--1940}.
\newblock


\bibitem[\protect\citeauthoryear{Kov{\'a}cs, Zucker, and Mazeh}{Kov{\'a}cs
  et~al\mbox{.}}{2002}]%
        {kovacs2002box}
\bibfield{author}{\bibinfo{person}{Geza Kov{\'a}cs}, \bibinfo{person}{Shay
  Zucker}, {and} \bibinfo{person}{Tsevi Mazeh}.}
  \bibinfo{year}{2002}\natexlab{}.
\newblock \showarticletitle{A box-fitting algorithm in the search for periodic
  transits}.
\newblock \bibinfo{journal}{\emph{Astronomy \& Astrophysics}}
  \bibinfo{volume}{391}, \bibinfo{number}{1} (\bibinfo{year}{2002}),
  \bibinfo{pages}{369--377}.
\newblock


\bibitem[\protect\citeauthoryear{Kumar, McCann, Naughton, and Patel}{Kumar
  et~al\mbox{.}}{2015}]%
        {KumarMNP15}
\bibfield{author}{\bibinfo{person}{Arun Kumar}, \bibinfo{person}{Robert
  McCann}, \bibinfo{person}{Jeffrey~F. Naughton}, {and}
  \bibinfo{person}{Jignesh~M. Patel}.} \bibinfo{year}{2015}\natexlab{}.
\newblock \showarticletitle{Model Selection Management Systems: The Next
  Frontier of Advanced Analytics}.
\newblock \bibinfo{journal}{\emph{{SIGMOD} Record}} \bibinfo{volume}{44},
  \bibinfo{number}{4} (\bibinfo{year}{2015}), \bibinfo{pages}{17--22}.
\newblock


\bibitem[\protect\citeauthoryear{Lazaridis and Mehrotra}{Lazaridis and
  Mehrotra}{2001}]%
        {LazaridisM01}
\bibfield{author}{\bibinfo{person}{Iosif Lazaridis} {and}
  \bibinfo{person}{Sharad Mehrotra}.} \bibinfo{year}{2001}\natexlab{}.
\newblock \showarticletitle{Progressive Approximate Aggregate Queries with a
  Multi-Resolution Tree Structure}. In \bibinfo{booktitle}{\emph{SIGMOD}}.
  \bibinfo{pages}{401--412}.
\newblock


\bibitem[\protect\citeauthoryear{Lazaridis and Mehrotra}{Lazaridis and
  Mehrotra}{2003}]%
        {conf/icde/LazaridisM03}
\bibfield{author}{\bibinfo{person}{Iosif Lazaridis} {and}
  \bibinfo{person}{Sharad Mehrotra}.} \bibinfo{year}{2003}\natexlab{}.
\newblock \showarticletitle{Capturing Sensor-Generated Time Series with Quality
  Guarantees}. In \bibinfo{booktitle}{\emph{Proceedings of the 19th
  International Conference on Data Engineering, March 5-8, 2003, Bangalore,
  India}}. \bibinfo{pages}{429--440}.
\newblock


\bibitem[\protect\citeauthoryear{Li, Yu, and Castelli}{Li
  et~al\mbox{.}}{1998}]%
        {conf/cikm/LiYC98}
\bibfield{author}{\bibinfo{person}{Chung{-}Sheng Li},
  \bibinfo{person}{Philip~S. Yu}, {and} \bibinfo{person}{Vittorio Castelli}.}
  \bibinfo{year}{1998}\natexlab{}.
\newblock \showarticletitle{{MALM:} {A} Framework for Mining Sequence Database
  at Multiple Abstraction Levels}. In \bibinfo{booktitle}{\emph{CIKM}}.
  \bibinfo{pages}{267--272}.
\newblock


\bibitem[\protect\citeauthoryear{Manku, Rajagopalan, and Lindsay}{Manku
  et~al\mbox{.}}{1998}]%
        {DBLP:conf/sigmod/RajagopalanML98}
\bibfield{author}{\bibinfo{person}{Gurmeet~Singh Manku},
  \bibinfo{person}{Sridhar Rajagopalan}, {and} \bibinfo{person}{Bruce~G.
  Lindsay}.} \bibinfo{year}{1998}\natexlab{}.
\newblock \showarticletitle{Approximate Medians and other Quantiles in One Pass
  and with Limited Memory}. In \bibinfo{booktitle}{\emph{{SIGMOD}}}.
  \bibinfo{pages}{426--435}.
\newblock


\bibitem[\protect\citeauthoryear{Mei and Moura}{Mei and Moura}{2017}]%
        {MeiM17}
\bibfield{author}{\bibinfo{person}{Jonathan Mei} {and}
  \bibinfo{person}{Jos{\'{e}} M.~F. Moura}.} \bibinfo{year}{2017}\natexlab{}.
\newblock \showarticletitle{Signal Processing on Graphs: Causal Modeling of
  Unstructured Data}.
\newblock \bibinfo{journal}{\emph{{IEEE} Trans. Signal Processing}}
  \bibinfo{volume}{65}, \bibinfo{number}{8} (\bibinfo{year}{2017}),
  \bibinfo{pages}{2077--2092}.
\newblock


\bibitem[\protect\citeauthoryear{Morse and Patel}{Morse and Patel}{2007}]%
        {DBLP:conf/sigmod/MorseP07}
\bibfield{author}{\bibinfo{person}{Michael~D. Morse} {and}
  \bibinfo{person}{Jignesh~M. Patel}.} \bibinfo{year}{2007}\natexlab{}.
\newblock \showarticletitle{An efficient and accurate method for evaluating
  time series similarity}. In \bibinfo{booktitle}{\emph{{SIGMOD}}}.
  \bibinfo{pages}{569--580}.
\newblock


\bibitem[\protect\citeauthoryear{Nelson}{Nelson}{1973}]%
        {nelson1973probability}
\bibfield{author}{\bibinfo{person}{Edward Nelson}.}
  \bibinfo{year}{1973}\natexlab{}.
\newblock \showarticletitle{Probability theory and Euclidean field theory}.
\newblock In \bibinfo{booktitle}{\emph{Constructive quantum field theory}}.
  \bibinfo{publisher}{Springer}, \bibinfo{pages}{94--124}.
\newblock


\bibitem[\protect\citeauthoryear{Pan, Hu, and Cao}{Pan et~al\mbox{.}}{2017}]%
        {pan2017construction}
\bibfield{author}{\bibinfo{person}{Zhuokun Pan}, \bibinfo{person}{Yueming Hu},
  {and} \bibinfo{person}{Bin Cao}.} \bibinfo{year}{2017}\natexlab{}.
\newblock \showarticletitle{Construction of smooth daily remote sensing time
  series data: a higher spatiotemporal resolution perspective}.
\newblock \bibinfo{journal}{\emph{Open Geospatial Data, Software and
  Standards}} \bibinfo{volume}{2}, \bibinfo{number}{1} (\bibinfo{year}{2017}),
  \bibinfo{pages}{25}.
\newblock


\bibitem[\protect\citeauthoryear{Pansare, Borkar, Jermaine, and Condie}{Pansare
  et~al\mbox{.}}{2011}]%
        {PansareBJC11}
\bibfield{author}{\bibinfo{person}{Niketan Pansare},
  \bibinfo{person}{Vinayak~R. Borkar}, \bibinfo{person}{Chris Jermaine}, {and}
  \bibinfo{person}{Tyson Condie}.} \bibinfo{year}{2011}\natexlab{}.
\newblock \showarticletitle{Online Aggregation for Large MapReduce Jobs}.
\newblock \bibinfo{journal}{\emph{{PVLDB}}} \bibinfo{volume}{4},
  \bibinfo{number}{11} (\bibinfo{year}{2011}), \bibinfo{pages}{1135--1145}.
\newblock


\bibitem[\protect\citeauthoryear{Park, Lee, and Chu}{Park
  et~al\mbox{.}}{1999}]%
        {park1999fast}
\bibfield{author}{\bibinfo{person}{Sanghyun Park}, \bibinfo{person}{Dongwon
  Lee}, {and} \bibinfo{person}{Wesley~W Chu}.} \bibinfo{year}{1999}\natexlab{}.
\newblock \showarticletitle{Fast retrieval of similar subsequences in long
  sequence databases}. In \bibinfo{booktitle}{\emph{KDEX}}.
  \bibinfo{pages}{60--67}.
\newblock


\bibitem[\protect\citeauthoryear{Philo}{Philo}{1997}]%
        {philo1997improved}
\bibfield{author}{\bibinfo{person}{John~S Philo}.}
  \bibinfo{year}{1997}\natexlab{}.
\newblock \showarticletitle{An improved function for fitting sedimentation
  velocity data for low-molecular-weight solutes}.
\newblock \bibinfo{journal}{\emph{Biophysical Journal}} \bibinfo{volume}{72},
  \bibinfo{number}{1} (\bibinfo{year}{1997}), \bibinfo{pages}{435--444}.
\newblock


\bibitem[\protect\citeauthoryear{Piatetsky{-}Shapiro and
  Connell}{Piatetsky{-}Shapiro and Connell}{1984}]%
        {ShapiroC84}
\bibfield{author}{\bibinfo{person}{Gregory Piatetsky{-}Shapiro} {and}
  \bibinfo{person}{Charles Connell}.} \bibinfo{year}{1984}\natexlab{}.
\newblock \showarticletitle{Accurate Estimation of the Number of Tuples
  Satisfying a Condition}. In \bibinfo{booktitle}{\emph{SIGMOD}}.
  \bibinfo{pages}{256--276}.
\newblock


\bibitem[\protect\citeauthoryear{Poosala, Ioannidis, Haas, and Shekita}{Poosala
  et~al\mbox{.}}{1996}]%
        {PoosalaIHS96}
\bibfield{author}{\bibinfo{person}{Viswanath Poosala},
  \bibinfo{person}{Yannis~E. Ioannidis}, \bibinfo{person}{Peter~J. Haas}, {and}
  \bibinfo{person}{Eugene~J. Shekita}.} \bibinfo{year}{1996}\natexlab{}.
\newblock \showarticletitle{Improved Histograms for Selectivity Estimation of
  Range Predicates}. In \bibinfo{booktitle}{\emph{SIGMOD}}.
  \bibinfo{pages}{294--305}.
\newblock


\bibitem[\protect\citeauthoryear{Potti and Patel}{Potti and Patel}{2015}]%
        {PottiP15}
\bibfield{author}{\bibinfo{person}{Navneet Potti} {and}
  \bibinfo{person}{Jignesh~M. Patel}.} \bibinfo{year}{2015}\natexlab{}.
\newblock \showarticletitle{{DAQ:} {A} New Paradigm for Approximate Query
  Processing}.
\newblock \bibinfo{journal}{\emph{{PVLDB}}} \bibinfo{volume}{8},
  \bibinfo{number}{9} (\bibinfo{year}{2015}), \bibinfo{pages}{898--909}.
\newblock


\bibitem[\protect\citeauthoryear{Reiss, Garofalakis, and Hellerstein}{Reiss
  et~al\mbox{.}}{2006}]%
        {ReissGH06}
\bibfield{author}{\bibinfo{person}{Frederick Reiss}, \bibinfo{person}{Minos~N.
  Garofalakis}, {and} \bibinfo{person}{Joseph~M. Hellerstein}.}
  \bibinfo{year}{2006}\natexlab{}.
\newblock \showarticletitle{Compact Histograms for Hierarchical Identifiers}.
  In \bibinfo{booktitle}{\emph{VLDB}}. \bibinfo{pages}{870--881}.
\newblock


\bibitem[\protect\citeauthoryear{Sidirourgos, Kersten, and Boncz}{Sidirourgos
  et~al\mbox{.}}{2011}]%
        {SidirourgosKB11}
\bibfield{author}{\bibinfo{person}{Lefteris Sidirourgos},
  \bibinfo{person}{Martin~L. Kersten}, {and} \bibinfo{person}{Peter~A. Boncz}.}
  \bibinfo{year}{2011}\natexlab{}.
\newblock \showarticletitle{SciBORQ: Scientific data management with Bounds On
  Runtime and Quality}. In \bibinfo{booktitle}{\emph{CIDR}}.
  \bibinfo{pages}{296--301}.
\newblock


\bibitem[\protect\citeauthoryear{Tobita}{Tobita}{2016}]%
        {tobita2016combined}
\bibfield{author}{\bibinfo{person}{Mikio Tobita}.}
  \bibinfo{year}{2016}\natexlab{}.
\newblock \showarticletitle{Combined logarithmic and exponential function model
  for fitting postseismic GNSS time series after 2011 Tohoku-Oki earthquake}.
\newblock \bibinfo{journal}{\emph{Earth, Planets and Space}}
  \bibinfo{volume}{68}, \bibinfo{number}{1} (\bibinfo{year}{2016}),
  \bibinfo{pages}{41}.
\newblock


\bibitem[\protect\citeauthoryear{Vlachos, Gunopulos, and Kollios}{Vlachos
  et~al\mbox{.}}{2002}]%
        {DBLP:conf/icde/VlachosGK02}
\bibfield{author}{\bibinfo{person}{Michail Vlachos}, \bibinfo{person}{Dimitrios
  Gunopulos}, {and} \bibinfo{person}{George Kollios}.}
  \bibinfo{year}{2002}\natexlab{}.
\newblock \showarticletitle{Discovering Similar Multidimensional Trajectories}.
  In \bibinfo{booktitle}{\emph{{ICDE}}}. \bibinfo{pages}{673--684}.
\newblock


\bibitem[\protect\citeauthoryear{Wang and Sevcik}{Wang and Sevcik}{2008}]%
        {WangS08}
\bibfield{author}{\bibinfo{person}{Hai Wang} {and} \bibinfo{person}{Kenneth~C.
  Sevcik}.} \bibinfo{year}{2008}\natexlab{}.
\newblock \showarticletitle{Histograms based on the minimum description length
  principle}.
\newblock \bibinfo{journal}{\emph{{VLDB J}.}} \bibinfo{volume}{17},
  \bibinfo{number}{3} (\bibinfo{year}{2008}), \bibinfo{pages}{419--442}.
\newblock


\bibitem[\protect\citeauthoryear{Wiscombe and Evans}{Wiscombe and
  Evans}{1977}]%
        {wiscombe1977exponential}
\bibfield{author}{\bibinfo{person}{WJ Wiscombe} {and} \bibinfo{person}{JW
  Evans}.} \bibinfo{year}{1977}\natexlab{}.
\newblock \showarticletitle{Exponential-sum fitting of radiative transmission
  functions}.
\newblock \bibinfo{journal}{\emph{J. Comput. Phys.}} \bibinfo{volume}{24},
  \bibinfo{number}{4} (\bibinfo{year}{1977}), \bibinfo{pages}{416--444}.
\newblock


\end{thebibliography}

\appendix

\section{Segmentation Algorithm}
\label{appendix:segmentation_alg}
We summarize the state-of-the-art time series segmentation algorithms, which can be classified into two categories: (i) Fix-length segmentation (\textsf{FL}), which partitions a time series based on fixed time windows.  The segments produced by the \textsf{FL} have equal lengths, and will be utilized in our aligned-segments experiments; and (ii)  Variable-length segmentation. There are three groups of algorithms produce variable-length segmentations: the \textsf{Top-down} methods~\cite{conf/cikm/LiYC98,park1999fast},  the \textsf{Bottom-up} approaches~\cite{conf/kdd/KeoghP98,keogh1999relevance} and  the \textsf{Sliding-window} techniques~\cite{koski1995syntactic,keogh2001online}. Among them, the \textsf{Sliding-window} (\textsf{SW}) has been proven to be more efficient than the \textsf{Top-down} and the \textsf{Bottom-up} methods~\cite{keogh2001online,keogh2004segmenting}. Thus, we choose the \textsf{Sliding-window (SW)} as the representative variable length segmentation algorithm in our experiments. The segments created by the \textsf{SW} have variable lengths~\cite{keogh2004segmenting} and are used in our misaligned-segments experiments. Figure~\ref{fig:intro_time} adopts the SW method, which produces variable-length segments.

\section{Propagating error guarantees in Arithmetic operators}
\label{appendix:propagate_errors}
For arithmetic operator $Ar_1\otimes\ Ar_2$ where $\otimes \{+, -, \times, \div\}$. If both $Ar_1$ and $Ar_2$ are scalar values, the \db{} gives accurate answers. Then we discuss in the following two cases: (i)  $Ar_1$ or $Ar_2$ is an aggregation result produced by \db, and (ii) both $Ar_1$ and $Ar_2$ are aggregation results produced by \db.

\noindent\textbf{Case 1}. Without loss of generality, we assume $Ar_1$ is an aggregation operator and $Ar_2$ is a scalar value. Let $\hat{R}$ be the approximate answer provided by \db{} for $Ar_1$ and $\hat{\varepsilon}$ is the corresponding error guarantee. The approximate answer and the error guarantee of  $Ar_1\otimes\ Ar_2$ is summarized in Table~\ref{fig:error_guarantees_general_case_1}.

\noindent\textbf{Case 2}. Both $Ar_1$ and $Ar_2$ are aggregation operators. Let $\hat{R}_1$ (resp. $\hat{R}_2$) and $\hat{\varepsilon}_1$ (resp. $\hat{\varepsilon}_2$) be the approximate answer and error guarantee provided by \db{} for $Ar_1$ and $Ar_2$ respectively. The approximate answer and the error guarantee of  $Ar_1\otimes\ Ar_2$ is summarized in Table~\ref{fig:error_guarantees_general_case_2}.

\section{Error guarantees of other expressions}
\label{sec:error_guarantee_others}
In this part, we present the error guarantees for the other core expressions, i.e., (i) \textsf{Sum}(\ser($v,a,b$)), (ii) \textsf{Sum}(Shift($T,k$)), (iii) \textsf{Sum}($T_1+T_2$), and (iv) \textsf{Sum}($T_1-T_2$).

\smallskip
\noindent\textbf{Error guarantee of \textsf{Sum}(\ser($v,a,b$)).}
For the time series $T$=\textsf{\ser}$(\upsilon, a, b)$, the estimation function is $\ef{T} = \upsilon$~\footnote{Under the reasonable assumption that any practical family will also include the constant function.}, then the error measures stored by \db{} are 
($\fes{T} = 0$, $\ses{T}=\upsilon\sqrt{b-a+1}$, $\tes{T}=0$). The error guarantee of \textsf{Sum}(\ser($v,a,b$)) is  $\tes{T}=0$.

\smallskip
\noindent\textbf{Error guarantee of \textsf{Sum}(Shift($T,k$)).}
For the time series $T$=\textsf{Shift}$(T, k)$, we need to use the error measures ($\fes{T}$, $\ses{T}$, $\tes{T}$) defined in domain $[a+k,b+k]$. Then the error guarantee of \textsf{Sum}(Shift($T,k$)) is  $\tes{T}$.

\smallskip
\noindent\textbf{Error guarantees of \textsf{Sum}($T_1+T_2$) and \textsf{Sum}($T_1-T_2$).}
Given two time series $T_1=(T_1^1,...,T_1^{k_1})$ and $T_2=(T_2^1,...,T_2^{k_2})$. Then the error measures of $T=T_1+ T_2$ are ($\fes{T}$, $\ses{T}$, $\tes{T}$) where  $\fes{T} = \sum_{i}^{k_1}\fes{T_1^i}+\sum_{i}^{k_2}\fes{T_2^i}$, $\ses{T} = \sum_{i}^{k_1}\ses{T_1^i}+\sum_{i}^{k_2}\ses{T_2^i}$, and $\tes{T} = \sum_{i}^{k_1}\tes{T_1^i}+\sum_{i}^{k_2}\tes{T_2^i}$. And the error guarantees of \textsf{Sum}($T_1+T_2$) is $\tes{T} = \sum_{i}^{k_1}\tes{T_1^i}+\sum_{i}^{k_2}\tes{T_2^i}$. The error measures of $T_1-T_2$ are the same with those of $T_1+T_2$.

\begin{figure}[h!]
\center
\renewcommand{\tabcolsep}{5mm}
\begin{tabular}{|l||c||c|} \hline
\textbf{Operator}                 &\textbf{approximate} &\textbf{error}      \\
				                  &\textbf{answer}      &\textbf{guarantee}  \\\hline
\textit{$Ar_1+ Ar_2$}             & $\hat{R} + Ar_2$                         &  $\hat{\varepsilon}$ \\\hline
\textit{$Ar_1- Ar_2$}             & $\hat{R} - Ar_2$                         &  $\hat{\varepsilon}$ \\\hline
\textit{$Ar_1\times Ar_2$}        & $\hat{R} \times Ar_2$                    &  $\hat{\varepsilon}\times Ar_2$ \\\hline
\textit{$Ar_1\div Ar_2$}          & $\hat{R} \div Ar_2$                      &  $\hat{\varepsilon}\div Ar_2$ \\\hline
\end{tabular}
\caption{Error guarantee propagation in case 1.}
\label{fig:error_guarantees_general_case_1}
\center
\renewcommand{\tabcolsep}{3mm}
\begin{tabular}{|l||c||c|} \hline
\textbf{Operator}                 & \textbf{approximate} & \textbf{error}      \\
				                  & \textbf{answer}      & \textbf{guarantee}  \\\hline
\textit{$Ar_1+ Ar_2$}             & $\hat{R}_1 + \hat{R}_2$                  &  $\hat{\varepsilon}_1+\hat{\varepsilon}_2$ \\\hline
\textit{$Ar_1- Ar_2$}             & $\hat{R}_1 - \hat{R}_2$                  &  $\hat{\varepsilon}_1+\hat{\varepsilon}_2$ \\\hline
\textit{$Ar_1\times Ar_2$}        & $\hat{R}_1 \times \hat{R}_2$             &  $\hat{\varepsilon}_1\hat{R}_2+ \hat{\varepsilon}_2\hat{R}_1+ \hat{R}_1\hat{R}_2$ \\\hline
\textit{$Ar_1\div Ar_2$}          & $\hat{R}_1 \div \hat{R}_2$               &  $\dfrac{(\hat{\varepsilon}_1\hat{R}_2+ \hat{\varepsilon}_2\hat{R}_1)}{(\hat{R}_2-\hat{\varepsilon}_2)\hat{R}_2}$ \\\hline
\end{tabular}
\caption{Error guarantee propagation in case 2.}
\label{fig:error_guarantees_general_case_2}
\end{figure}

\section{Computation of Formula~\ref{equ:aligned_ANY}}
\label{appendix:aligned_ANY}
\begin{align*}
\varepsilon &= \Big|\sum_{i=a}^{b}\ts_1[i]\ts_2[i] - \sum_{i=a}^{b}\ef{T_1}(i)\ef{T_2}(i)\Big|\nonumber\\
& = \Big|\sum_{i=1}^{k}\Big(\sum_{j=a_i}^{b_i}\ts_1[i]\ts_2[i] - \sum_{j=a_i}^{b_i}\ef{T_1}(i)\ef{T_2}(i)\Big)\Big|\nonumber\\
& = \Big|\sum_{i=1}^{k}\Big(\langle \bm{\varepsilon}_{T_1^i}, \ef{T_2^i} \rangle +  \langle \bm{\varepsilon}_{T_2^i}, \ef{T_1^i} \rangle + \langle \bm{\varepsilon}_{T_1^i}, \bm{\varepsilon}_{T_2^i} \rangle \Big) \Big|\nonumber\\
& \leq \Big|\sum_{i=1}^{k}\langle \bm{\varepsilon}_{T_1^i}, \ef{T_2^i} \rangle  \Big|+  \Big|\sum_{i=1}^{k}\langle \bm{\varepsilon}_{T_2^i}, \ef{T_1^i} \rangle  \Big|+ \Big|\sum_{i=1}^{k}\langle \bm{\varepsilon}_{T_1^i}, \bm{\varepsilon}_{T_2^i} \rangle  \Big|\nonumber\\
&\leq \sum_{i=1}^{k}\Big(\fes{T_1^i}\fes{T_2^i}+\fes{T_1^i}\ses{T_2^i} + \ses{T_1^i}\fes{T_2^i}\Big)
\end{align*}
The last inequality is obtained by Applying the H$\ddot{o}$lder inequality~\cite{cheney2009linear}.

{\small
\begin{table}[t]
\centering
\renewcommand{\tabcolsep}{1mm}
\begin{tabular}{|c|c|c|}\hline
$\tau$ & Expression & Comment\\ \hline \hline 
$p_i$  & $\{\sum_{i=0}^i a_ix^i |a_i\in R\}$ & i-degree Polynomial\\\hline
$g$    & $\{a\exp(\frac{-(x-b)^2}{2c^2})+d|a,b,c,d\in R\}$ & Gaussian\\\hline
$l$    & $\{\frac{L}{1+\exp(ax+b)}+c|L,a,b\in R\}$ & Logistic\\\hline
\end{tabular}
\caption{Example function family identifiers}
\label{table:funciton_family_identifier}
\renewcommand{\tabcolsep}{0.5mm}
\centering
\begin{tabular}{|c|c|c|c|} \hline
				 & \multicolumn{3}{c|}{Generated Error Measures} \\\hline
                 & $\fes{T}$               &  $\ses{T}$               & $\tes{T}$            \\\hline \hline 
$\ts = \plusOP$  & $\fes{T_1}+\fes{T_2}$   &  $\ses{T_1}+\ses{T_2}$   & $\tes{T_1}+\tes{T_2}$\\\hline  
$\ts = \minusOP$ & $\fes{T_1}+\fes{T_2}$   &  $\ses{T_1}+\ses{T_2}$   & $\tes{T_1}+\tes{T_2}$\\\hline  
\multirow{4}{*}{$\ts = \timesOP$} &   $\fes{T_1}\fes{T_2}$   &  \multirow{4}{*}{$\ses{T_1}\ses{T_2}$}   & \cellcolor{gray!15}$\fes{T_1}\fes{T_2}$\\\cline{4-4}
                                  &   $+\fes{T_1}\ses{T_2}$   &                                         &   $\fes{T_1}\fes{T_2}$   \\
                                  &   $+\fes{T_2}\ses{T_2}$   &                                         &   $+\fes{T_1}\ses{T_2}$  \\
                                  &                          &                                          &   $+\fes{T_2}\ses{T_2}$   \\\hline
\end{tabular}
\caption{Error measures propagation. $\tes{T=\timesOP}$ has two possible computation methods. If the estimation function family forms a vector space, then we use the one in the grey background.}
\label{tbl:error-guarantee-propagation}
\end{table}
} \normalsize

\section{Proof of the optimality of OS}
\label{appendix:proof_optimal_segment_combination}
\eat{
\begin{proof}
As Algorithm~\ref{alg:optimal_segmentation}  chooses the segment combination based on the values of $E_{T_1}$ and $E_{T_2}$ where $E_{T_1}$ and $E_{T_2}$ are all the possible error guarantee combinations of the segments in $T_1$ and $T_2$ respectively. The segment combination returned by Algorithm~\ref{alg:optimal_segmentation}  is the one leading to the minimal combined values from $E_{T_1}$ and $E_{T_2}$. Thus, there do not exist other segment combinations with smaller error guarantees, which proves the optimality of Algorithm~\ref{alg:optimal_segmentation}.
\end{proof}
}

\begin{proof}
We use a proof by induction to show that the error guarantee produced by the segment combination returned by OS (Algorithm~\ref{alg:optimal_segmentation}) is  optimal.

Let $OPT(\rep{T_1},\rep{T_2})= \{[a_i,b_i] | i\in[1,m]\}$ be the segment combination returned by OS. First, let's see the base case where $OPT(\rep{T_1},\rep{T_2})= \{[a_1,b_1]\}$ has only one domain. There are two cases depending on $b_1=b_1^1$ or $b_1=b_2^t$ where $\cover{T_2}{[a_1,b_1]}= \{T_2^1,...,T_2^t\}$. 

Case 1: $b_1 = b_1^1$. Since OS chooses $[a_1, b_1^1]$ as the domain, then $b_2^1\leq b_1^1$. Otherwise, OS does not choose $[a_1, b_1^1]$. This is because, (i) if $E_{T_1}[0]\geq E_{T_2}[0]$ then OS will choose $[a_1, b_2^1]$ instead; or (ii) if $E_{T_1}[0]< E_{T_2}[0]$, then OS can not enter the loop in lines 8 - 17. Since $b_2^1\leq b_1^1$, then we know $E_{T_1}[0]\leq E_{T_2}[0]$, so the error guarantee is $\fes{T_1^1}\fes{T_2^1}$, which is the minimal error guarantee in domain $[a_1, b_1]$. Assume we split the domain $[a_1,b_1]$ into $p$ ($p\geq 2$) sub-domains $[a_1, c_1], [c_1,c_2],...,$ $ [c_{p-1},b_1]$, then the error guarantee is $p\fes{T_1^1}\fes{T_2^1}$, therefore, domain $[a_1, b_1]=[a_1^1, b_1^1]$ produces the minimal error guarantee.

Case 2: $b_1=b_2^t$. Since OS chooses $[a_1, b_2^t]$ as the domain, we know that the error guarantee is  
\begin{align*}
(\sum_{i\in\cover{T_2}{[a_2^1,b_2^t]}}\fes{T_2^1}^2)^{\frac{1}{2}}(\sum_{i\in\cover{T_1}{[a_2^1,b_2^t]}}\fes{T_1^1}^2)^{\frac{1}{2}}
\end{align*} 
which is less than $\fes{T_1^1}(\sum_{i\in\cover{T_2}{[a_1^1,b_1^1]}}\fes{T_2^1}^2)^{\frac{1}{2}}$. If we split $[a_2^1, b_2^t]$ into several sub-domains, the error guarantee is greater than $\fes{T_1^1}(\sum_{i\in\cover{T_2}{[a_1^1,b_1^1]}}\fes{T_2^1}^2)^{\frac{1}{2}}$. Thus, $[a_1, b_1]=[a_2^1, b_2^t]$ produces the minimal error guarantee. 

Suppose $OPT(\rep{T_1},\rep{T_2})= \{[a_i,b_i] | i\in[1,m-1]\}$ produces the minimal error guarantee, then for the case   $OPT(\rep{T_1},\rep{T_2})= \{[a_i,b_i] | i\in[1,m]\}$, we only need to prove the last domain $[a_m, b_m]$ produces the minimal error guarantee, which is the same to the base case.
\end{proof}

\section{Proof of Lemma~1}
\label{appendix:proof_of_polynomial_LSF}
\begin{proof}
Let $\ff=\{\sum_i \alpha_i t^i$ |$\alpha_i \in R\}$ be a polynomial function family defined on $[a,b]$. The restriction of $f\in \ff$ on $[a', b']\subseteq [a,b]$ is $f|_{[a',b']}$ = $(a', b', [\sum_i \alpha_i (a')^i,$ $...,\sum_i \alpha_i (b')^i])$. The shift of $f|_{[a',b']}$ to $a-a'$ steps is \textsf{Shift}$(f|_{[a',b']}, a-a')$ = $(a, a +b' - a', [\sum_i \alpha_i (a')^i, ...,$ $\sum_i \alpha_i (b')^i])$. $[\sum_i \alpha_i (a')^i, ...,$ $\sum_i \alpha_i (b')^i]$ can be transformed into $[\sum_i \beta_i (a)^i, ...,$ $\sum_i \beta_i (a+b'-a')^i]$ such that $\beta_i =\frac{\alpha_i(a'+k)^i}{(a+k)^k}$ for all $i\in[a, a+b'- a']$. Let $f' = \sum_i \beta_i t^i$ be a function in $\ff$. Thus $f'|_{[a, a+b'-a']} =  [\sum_i \beta_i (a)^i, ...,$ $\sum_i \beta_i (a+b'-a')^i] = $ $Shift(f|_{[a',b']},a-a')$. 
\end{proof}

\section{Proof the correctness and tightness of Equation~5}
\label{appendix:proof_misaligned_times_LSF}
\begin{proof}
Let $\varepsilon_{Sum(T_1\times T_2)}$ be the true error of $Sum(T_1\times T_2)$.
\begin{align*}
\varepsilon_{Sum(T_1\times T_2)} &= |\langle \bm{\varepsilon}_{T_1}, \bm{f}_{T_2} \rangle + \langle \bm{\varepsilon}_{T_2}, \bm{f}_{T_1} \rangle + \langle \bm{\varepsilon}_{T_1}, \bm{\varepsilon}_{T_2} \rangle|\\
&\leq |\langle \bm{\varepsilon}_{T_1}, \bm{f}_{T_2} \rangle| + |\langle \bm{\varepsilon}_{T_2}, \bm{f}_{T_1} \rangle| + |\langle \bm{\varepsilon}_{T_1}, \bm{\varepsilon}_{T_2} \rangle|
\end{align*}

The first term $|\langle \bm{\varepsilon}_{T_1}, \bm{f}_{T_2} \rangle|$ can be rewritten as
\begin{align*}
&|\langle \bm{\varepsilon}_{T_1}, \bm{f}_{T_2} \rangle| = \big|\sum\limits_{i=1}^{k_1} \langle \bm{\varepsilon}_{T_1}|_{[a_i,b_i]}, \bm{f}_{T_2}|_{[a_i,b_i]} \rangle\big|\\
&= \Big|\sum\limits_{i=1}^{k_1} \Big(\overbrace{\langle \bm{\varepsilon}_{T_1}|_{[a_1^i,b_1^i]}, \ef{T_1^i} \rangle}^{=0} + \langle \bm{\varepsilon}_{T_1}|_{[a_1^i,b_1^i]}, \bm{f}_{T_2}|_{[a_1^i,b_1^i]}-\ef{T_1^i} \rangle\Big)\Big|\\
&\leq \sum\limits_{i=1}^{k_1} \Big|\Big(\langle \bm{\varepsilon}_{T_1}|_{[a_1^i,b_1^i]}, \bm{f}_{T_2}|_{[a_1^i,b_1^i]}-\ef{T_1^i} \rangle\Big)\Big|\\
&\leq \sum\limits_{i=1}^{k_1} \big\|\bm{\varepsilon}_{T_1}|_{[a_1^i,b_1^i]}\big\|_2 \big\|\bm{f}_{T_2}|_{[a_1^i,b_1^i]}-\ef{T_1^i}\big\|_2\\
&= \sum\limits_{i=1}^{k_1} \fes{T_1^i}\|\bm{f}_{T_2}|_{[a_1^i,b_1^i]}-\ef{T_1^i}\|_2
\end{align*}
\yannisp{I think that the orthogonal property is not sufficient to prove the next point. The LSF is also needed. This needs to be clarified and explain why the LSF is critical at this point. If I am wrong and the LSF dependency is not here but at some other point, explain which one is this other point. Chunbin: this is a comment for old version.}
%Notice that, $\langle \bm{\varepsilon}_{T_1}|_{[a_i,b_i]}, \hat{f}^{T_1^i}  \rangle = 0$ for all $i\in[1,k_1]$
%is guaranteed by the orthogonal property. Therefore, we have
Similarly, we have:
\begin{align*}
|\langle \bm{\varepsilon}_{T_2}, \bm{f}_{T_1} \rangle|  \leq \sum\limits_{i=1}^{k_2} \Big(\|\varepsilon_{T_2^i}\|_2 \times \|\bm{f}_{T_1}|_{[a_2^i,b_2^i]}-\ef{T_2^i}\|_2\Big)
\end{align*}
Recall that the computation of $|\langle \bm{\varepsilon}_{T_1}, \bm{\varepsilon}_{T_2}\rangle|$ is presented in Section~\ref{sec:segment_combination_selection}. Combining the results of $|\langle \bm{\varepsilon}_{T_1}, \bm{\varepsilon}_{T_2}\rangle|$, $|\langle \bm{\varepsilon}_{T_1}, \bm{f_{T_2}}\rangle|$, and $|\langle \bm{f_{T_1}}, \bm{\varepsilon}_{T_2}\rangle|$ completes the proof. 
\end{proof}

\section{Computation of Formula~5}
\label{appendix:computation_formula_5}
Here we present how to compute $\|\bm{f_{T_2}|_{[a_1^i,b_1^i]} - \ef{T_1^i}}\|_2$ and $\|\bm{f_{T_1}|_{[a_2^i,b_2^i]} - \ef{T_2^i}}\|_2$ in $O(dim(\ff)^3)$.

Let's first look into $\|\bm{f_{T_2}|_{[a_1^i,b_1^i]} - \ef{T_1^i}}\|_2$.

\begin{align*}
&\|\bm{f_{T_2}|_{[a_1^i,b_1^i]} - \ef{T_1^i}}\|_2 = \\
&\Big(\sum_{j\in \cover{T_2}{[a_1^i,b_1^i]}} \|\bm{f_{T_2}|_{[a_1^i,b_1^i]\cap[a_2^j, b_2^j]}}-\bm{\ef{T_1^i}|_{[a_1^i,b_1^i]\cap[a_2^j, b_2^j]}}\|_2^2\Big)^{\frac{1}{2}}\\
&=\Big(\sum_{j\in \cover{T_2}{[a_i,b_i]}} \|\Psi([a_2^i,b_2^i],[a_1^i,b_1^i]\cap[a_2^j, b_2^j])\ef{T_2^j}\\
&~~~~~~-\Psi([a_1^i,b_1^i],[a_1^i,b_1^i]\cap[a_2^j, b_2^j])\ef{T_1^i}\|_2^2\Big)^{\frac{1}{2}}
\end{align*}
where $\Psi$ is an orthonormal basis transformation matrix, which can be computed in $O(dim(\mathds{F})^3)$. $\Psi([a_2^i,b_2^i],[a_1^i,b_1^i]\cap[a_2^j, b_2^j])$ transforms the orthonormal basis from the domain $[a_2^i,b_2^i]$ to the sub-domain $[a_1^i,b_1^i]\cap[a_2^j, b_2^j]$. In the following, we will show the details of computing $\Psi$.

Given a function family $\mathds{F}$, let $(\varphi_i^{[a,b]})_{1\leq i \leq dim(\mathds{F})}$ be an orthonormal basis of $\mathds{F}$ on the domain $[a,b]$ for the scalar product $\langle f_1, f_2\rangle = \sum_{i=a}^b (f_1(i)\times f_2(i))$ where $f_1,f_2\in \mathds{F}$. Such orthonormal basis can be obtained by using the Gram$-$Schmidt process~\cite{hoffmann1989iterative}. Given a domain $[a,b]$ and one sub-domain  $[a',b']\subset [a,b]$, let $\Psi([a,b], [a',b'])$ be the basis transform matrix such that
\begin{align*}
\Psi([a,b], [a',b'])_{i,j} = \langle\varphi_i^{[a,b]}|_{[a',b']}, \varphi_j^{[a',b']}\rangle
\end{align*}
That is using $\Psi$, we can directly obtain the  orthonormal basis for any sub-domain. The size of $\Psi$ is $dim(\ff)^2$ and the computation of each $\Psi([a,b], [a',b'])_{i,j}$ is $O(dim(\ff))$. Therefore, the overall cost of computing $\Psi$ is $O(dim(\ff)^3)$.

\begin{figure}[t]
\center
\includegraphics[width=0.5\textwidth]{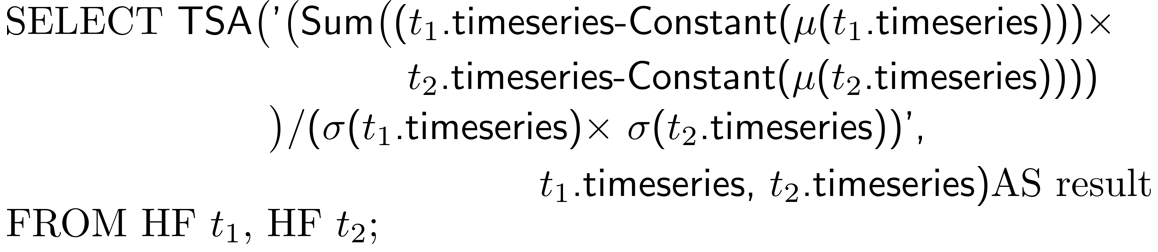}
\vspace{-5mm}
\caption{SQL query computing correlation TSA for all the time series pairs in HF.}
\label{fig:sql_query}
\end{figure}

\section{Experiment Setting Details}
\label{appendix:experiment}
\noindent\textbf{Datasets}.
We evaluated all the error guarantee methods on four real-life datasets.
\begin{compact_item}
  \item Historical Forex Data (HF) %\footnote{\url{https://pepperstone.com/en/client-resources/historical-tick-data}} 
   are tick-by-tick market data for $15$ Forex (foreign exchange) data pairs, e.g., AUD/JPY	(Australian Dollar vs. Japanese Yen) from May 2009 to November 2016. Each Forex pair is considered a time series with $\sim126$ million data points (3 per second).
\item Historical IoT Data (HI) were provided by Teradata and measure the internal oil pressure and the oil temperature every second from $8/19/2015$ to $11/17/2015$, as reported by seven engines in mining trucks in Chile.
\item Historical Bitcoin Exchanges Data (HB)\footnote{\url{https://www.kaggle.com/mczielinski/bitcoin-historical-data/data}} contains $16$ cryptocurrency exchange prices per minute from January 2012 to January 2018. Each cryptocurrency is considered as a time series.
\item Historical Air Quality Data (HA)\footnote{\url{https://www.kaggle.com/ktochylin}} present $11$ different air quality measurements such as air pressure, air temperature and relative humidity from $09/10/2011$ to $09/10/2014$ in San Diego, at 1-minute resolution.
%\url{https://www.kaggle.com/ktochylin/san-diego-every-minute-weather-indicators-201114/data}
%The HA data follows some obvious patterns and it has underlying logics.
%    \item Historical Climate Data (HC)\footnote{\url{http://www.met.ie/climate-request/}}: It stores the historical climate data, such as the air temperature, the  elative humidity, and the vapour pressure,  of $25$ stations in $15$ countries of Ireland from $01/01/1988$ to $10/31/2017$ every hour. For each station, we choose three time series, i.e.,  the air temperature, the  relative humidity, and the vapour pressure.
\end{compact_item}
%Table~\ref{tbl:data_characteristics} summarizes the data characteristics.

The HF and HB are financial market data, which are considered hard-to-model, while HI and HA are climate data following certain patterns. For example, the temperature in afternoon is usually higher than that at night, etc.
Not surprisingly, the HB experiments behaved very similarly to the HF experiments, while the HA experiments behaved similarly to the HI ones. 

\noindent\textbf{Queries}.
The SQL query computing the correlation TSA for all the time series pairs in HF is shown in Figure~\ref{fig:sql_query}. The SQL queries on the other three datasets are similar by change the table HF to HI, HB and HA respectively.

\end{document}